\documentclass[superscriptaddress,twocolumn,preprintnumbers,amsmath,amssymb,floatfix,aps,prb]{revtex4-2}

\usepackage[english]{babel}

\usepackage{graphicx,tabularx}
\usepackage[version=3]{mhchem} 
\usepackage{amssymb}
\usepackage{dcolumn}
\usepackage{color}
\usepackage[mathcal]{euscript}

\definecolor{gray0}{gray}{0.0}
\definecolor{gray64}{gray}{0.25}
\definecolor{gray128}{gray}{0.5}
\definecolor{gray192}{gray}{0.75}
\definecolor{gray255}{gray}{1.0}

\renewcommand{\thesection}{\arabic{section}}
\renewcommand{\thesubsection}{\thesection.\arabic{subsection}}
\makeatletter
\renewcommand{\p@subsection}{}
\makeatother

\newcommand{\review}[1]
{\color{black}{{#1}\ }\color{black}}

\begin{document}
\title{Modeling the Zero-Phonon Line of Strained SnV Centers in Diamond; Including Reflections on Computational Cost and Accuracy.}
\author{Danny E. P. Vanpoucke}
\email[Corresponding author: ]{Danny.Vanpoucke@UHasselt.be}
\affiliation{UHasselt, Institute for Materials Research (IUMAT), Quantum \& Artificial inTelligence design Of Materials (QuATOMs), Martelarenlaan 42, B-3500 Hasselt, Belgium}
\affiliation{imec, IUMAT, Wetenschapspark 1, B-3590 Diepenbeek, Belgium}

\begin{abstract}
Among the group-IV vacancy color centers in diamond, the SnV holds promise for photonics based quantum applications. In this work, the Tin-Vacancy (SnV) zero-phonon line (ZPL) and its pressure coefficient are calculated using first principles approaches. The predicted absolute ZPL position is shown to be strongly influenced by the method and supercell size used. The results are therefore extrapolated to the dilute limit allowing for direct comparison with experiments. The importance of identifying the color-center related Kohn--Sham states is highlighted, as well as the shifting of these states due to electron excitations as well as supercell size and k-point position. In contrast to the absolute ZPL positions, the relative position of the SnV$^0$ ZPL is consistently redshifted about $43$ nm compared to the SnV$^-$ ZPL. In addition, the pressure coefficient is shown to be very robust over different methods, always resulting in a value of about $1.4$ nm/GPa, for both SnV$^0$ and SnV$^-$. Finally, the computational accuracy and cost are put into perspective. 
\end{abstract}

\maketitle
\section{Introduction}\label{sec:intro}
In recent years, color centers in diamond have attracted increasing attention due to their potential application in quantum technologies as, for example, qubits or single photon emitters\cite{GaliA:2018npjCompMat}. Single photon emitters have their application in quantum information technology, where having a strong zero-phonon line (ZPL) in the telecom band is of central interest, as this reduces fiber loss\cite{RufM:2021JAppPhys,Mukherjee:2023NanoLett,Filippatos:2025JApplPhys}. Also in high resolution sensors with optical readout for biomedical, temperature, or pressure applications, the ZPL of these color centers is of interest due to their sensitivity to the external environment\cite{DamleVVanpouckeDEP:2020bCarbon,Schirhagl:2022AnalChem_revBiochem,Neumann:2013nanoLett_DiamTermo,VanWijkVanpouckeDEP:2025Carbon,Ekimov:2025PhysRevB_G4strainZPL}.\\
\indent Although several hundred color centers have been experimentally identified, only for a handful, the structural geometry has been established, with the Nitrogen--Vacancy (NV) and group-IV split-vacancy color centers the most well-known\cite{Zaitsev:2001ColorCenters,VanpouckeDEP:2025BookChapColor}. Where basic experimental characterization of the ZPL is useful from a cataloging perspective, for technological applications and engineering of the ZPL, the effective atomic structure is an essential piece of information. It allows for the predictive modeling of color center properties and responses, by means of quantum mechanical simulations. Even though the inverse problem of predicting the atomic structure associated with a given ZPL peak is currently intractable, accurate density functional theory (DFT) studies can help selecting among the proposed structural models\cite{GossJP:1996prl,GaliThiering:2021PhysRevRes_NiV}. In contrast, given an accepted atomic scale structure, DFT can be used to predict ground state configurations, the impact of pressure and strain, the interaction with other defects, etc\cite{GossJP:2005PRB, VanpouckeDEP:DiamRelatMater2017,LofgrenR:NewJPhys2018,lindner:2018njp,ekimovE:2019jetp,VanpouckeDEP:DiamRelatMater2019, VanWijkVanpouckeDEP:2025Carbon, Filippatos:2025JApplPhys}.\\
\indent Of all color centers in diamond, the NV$^-$ center is probably the most studied and the effective workhorse at the forefront of quantum research. However, due to its weak ZPL, with a Debye--Waller factor of only $0.03$\cite{DOHERTY:2013pr}, there is an active search for suitable alternatives. The group-IV color centers (SiV\cite{NeuE:2011prb}, GeV\cite{Siyushev:2017prb,iwasaki:2015SR}, SnV\cite{Iwasaki:2017PRL_SnV}, and PbV\cite{Wang:2024PhysRevLett_PbV,Abe:2026AdvFunctMat_PbV}) are considered promising candidates due to their bright ZPL and smaller phonon sideband. In addition, these split-vacancy color centers (effectively six-coordinated group-IV dopants with $D_{3d}$ symmetry)\cite{GossJP:2005PRB,thiering:2018prx, WahlPereira:PhysRevLett2020_SnV}, have an inversion symmetry protecting the optical transitions from first-order electric field fluctuations. This makes them attractive as potential quantum network nodes in photonic based platforms and other nanophotonic applications\cite{chen:2020aqt,RufM:2021JAppPhys}.\\ 
\indent The Tin-split-Vacancy (SnV) color center has attracted attention due to its long coherence time at temperatures below 2K, as well as its large ground state splitting of $850$ GHz\cite{Karapatzakis:2024PhysRevX_GSsplit,RufM:2021JAppPhys}. It has a ZPL at $620$ nm with a Debye-Waller factor of $0.57$, which has been supported by first principles calculations\cite{Iwasaki:2017PRL_SnV,MaryJoyVanpoucke:ACSMaterSci2026,Gorlitz:2020NJPhys,thiering:2018prx,ekimov:2018DRM}. In the work of Iwasaki \textit{et al.}\cite{Iwasaki:2017PRL_SnV} the same trends for the series SiV, GeV, and SnV were found from calculations, in agreement with earlier work of Goss \textit{et al.}\cite{GossJP:2005PRB}, with the ZPL calculated to be at about $1.6$ eV. Ekimov \textit{et al.}\cite{ekimov:2018DRM} using a non-conventional $83$ atom supercell report excitation energies of $1.888$ and $1.867$ eV for the SnV$^-$ and SnV$^0$ color center, though it is unclear which functional was used for their calculation. Thiering and Gali\cite{thiering:2018prx} have shown that the group-IV vacancy color centers are dynamic Jahn--Teller systems, and predict the energy contribution for the SnV color center to be of the order of $80$ meV, while spin-orbit contributions provide a compensating factor of about $20$ meV. This resulted in a ZPL prediction of $2.09$ eV for the SnV$^-$ color center, using a $512$ atom supercell and range-separated hybrid functional.\\
\indent In recent years significant experimental progress has been made to establish the SnV$^-$ as platform for quantum network nodes in scalable quantum technologies\cite{Brevoord:2025Thesis_SnVexp}. It has been shown that its optical transitions can be tuned through strain engineering \cite{Brevoord:2025ApplPhysLett,Li:2024Nature_SnVinCMOS}, however, the performance of the SnV$^-$ color center is also influenced by the same strain in the local environment\cite{Sedov:2023PhilTransSoc_SnV, Gorlitz:2020NJPhys}. G\"{o}rlitz \textit{et al}.\cite{Gorlitz:2020NJPhys} showed that at an anneal temperature of $2100\ ^\circ$C and pressure of $7.7$ GPa a clear double peaked ZPL-spectrum at $620$ nm is obtained, while annealing at $1200\ ^\circ$C and $10^{-6}$ mbar resulted in a spectrum containing a distribution of ZPL lines ranging from $612$ to $630$ nm, a result of localized strain due to lattice damage acting on individual color centers. Recent work by Mary Joy \textit{et al.}\cite{MaryJoyVanpoucke:ACSMaterSci2026} showed that annealing at $850 ^\circ$C compared to $750 ^\circ$C, reduces the number of active color centers and transforms the broad distribution of ZPL into a bimodal distribution separated about $6$ nm. \review{This is in line with previous observations of ion beam implanted SnV, showing a reduction of bond-centered Sn in favor of substitutional Sn after annealing at $920 ^\circ$C\cite{WahlPereira:PhysRevLett2020_SnV}.}  Furthermore, Sedov \textit{et al}.\cite{Sedov:2023PhilTransSoc_SnV} hypothesized that the ZPL shift was not only influenced by the absolute size of the local strain, but also the ratio of longitudinal and transversal strain. This is in agreement with the predictions by van Wijk \textit{et al.}\cite{VanWijkVanpouckeDEP:2025Carbon} for the GeV$^0$. They predict both red- and blueshifted ZPL, with the pressure coefficients being smaller for hydrostatic strain compared to linear strain. Vindolet \textit{et al.}\cite{VindoletB:PRB2022} observed for the SiV and GeV color centers a significant blueshift of the ZPL under hydrostatic compression. The effect at $180$ GPa was found to be about five times larger for the GeV than for the SiV color center. An even larger effect was predicted for the SnV. Corte \textit{et al.}\cite{Corte:2021AdvPhotonRes_SnV0} investigated the ZPL lines observed at $620$, $631$, and $647$ nm attributed to the SnV color center\cite{tchernij:2017acs}, with the first one being assigned to the SnV$^-$ center. However, the nature of the $647$ nm ZPL remains unclear, although initially tentatively attributed to the SnV$^0$ color center.\\
\indent Understanding the impact of strain, both intentional, in the case of strain engineering, or as a consequence of local lattice defects, is of importance for the application of SnV color centers in quantum optical frameworks and devices. In the current work, the ZPL and pressure coefficient of the SnV$^0$ and SnV$^-$ are investigated using density functional theory. Results obtained using different methodological approaches are compared to provide a guideline for best practices directed at using density functional theory as predictive tool. In addition to considering potential accuracy of the different approaches, the computational cost is also discussed to limited extent, as this is a factor often unfamiliar to the experimental researcher.\\
\section{Methodology}\label{sec:meth}
\subsection{Computational Details \label{subsec:compdet}}
First principles \review{spin-polarized} density functional theory calculations are performed using a projector augmented waves method as implemented in the Vienna Ab Initio Simulation Package (VASP.6.4.3)\cite{KresseG:PhysRevB1999,KresseG:PhysRevB1996}. The kinetic energy cutoff was set to $600$ eV, and the generalized gradient approximation (GGA) functional PBE by Perdew, Burke and, Ernzerhof, and range-separated hybrid functional HSE06 by Heyd, Scuseria and, Ernzerhof, are used to describe the exchange and correlation behavior of the valence electrons (C: $2s^22p^2$ and Sn: $4d^{10}5s^25p^2$)\cite{PBE_PerdewJP:PhysRevLett1996, HSE06paper_KrukauAV:JChemPhys2006}. The latest generation 64, GW-ready standard pseudo-potentials are used. Since the supercells are relatively large and optical excitations are vertical in nature, the first Brillouin-zone is sampled using the $\Gamma$-point only. Exceptions where an extended k-point sampling is used are clearly indicated and always include the $\Gamma$-point. A conjugate gradient method is used for structure optimization. Only atomic positions are optimized, while the lattice vectors are constrained to a cubic geometry. Equilibrium volumes for the system including the color centers are determined by fitting the Rose--Vinet equation of state to energy-volume data\cite{VinetP:PhysRevB1987_RoseVinetEOS}. The energy convergence criterion for the Self-Consistent Field (SCF) method is set to $1.0\times10^{-8}$, and at least $1.0\times10^{-4}$ for PBE and HSE06 calculations, respectively. After structure optimization, the maximum forces remaining on a single atom are below $1$ meV/\AA\ for PBE calculations, and below $30$ meV/\AA\ at the HSE06 level.\\

\begin{figure}[!tb]
    \centering
    \includegraphics[width=1.0\linewidth]{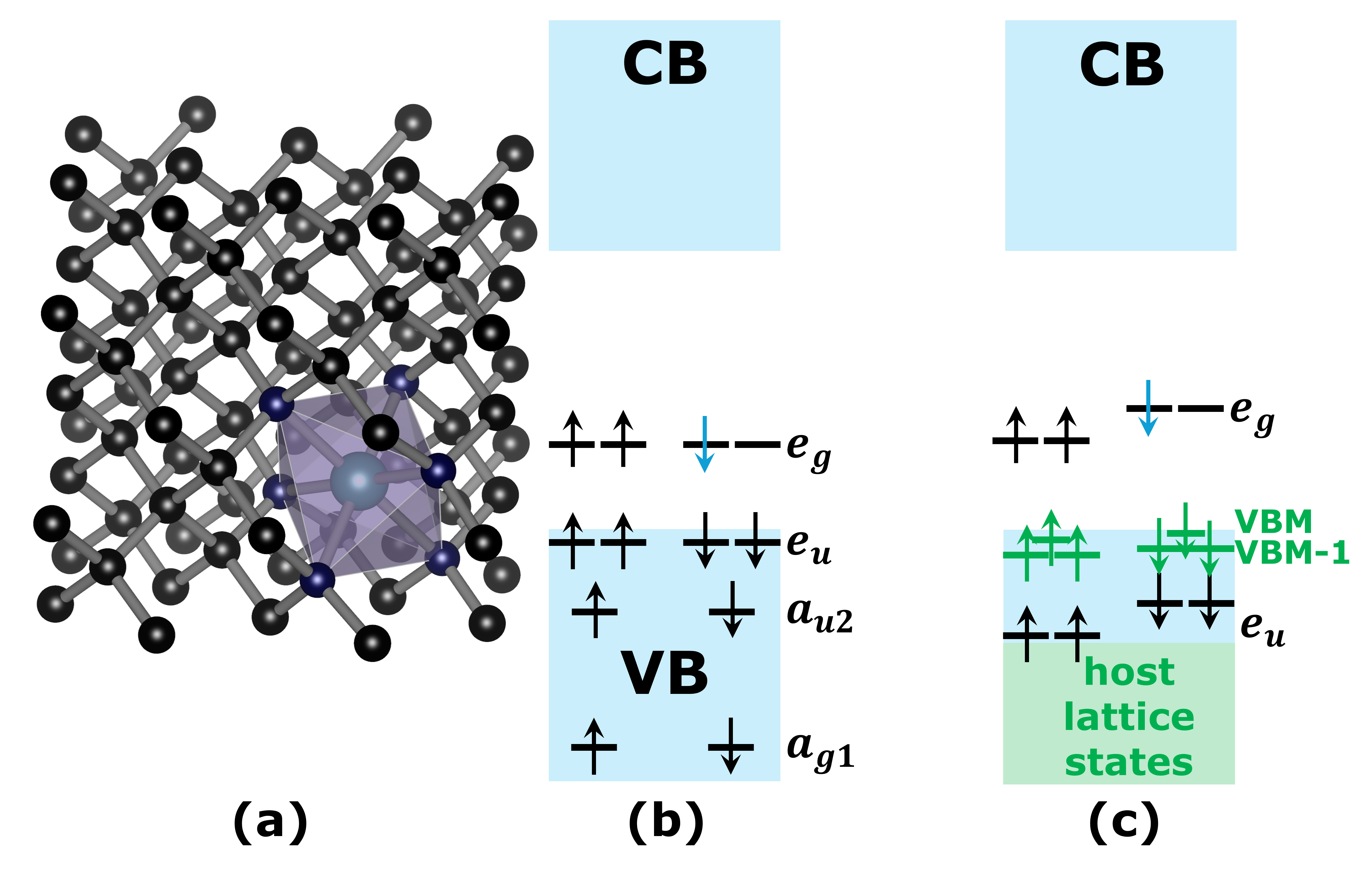}
    \caption{(a) Ball-and-stick representation of the local environment of the SnV color center. The Sn atom is shown in teal, and the six bonded C atoms are shown in dark blue. Together, these seven atoms are considered to represent the color center in diamond. (b) Schematic representation of the color center states according to group theory, and (c) their effective positioning as well as relevant states at the top of the valence band in practical DFT calculations.}
    \label{fig:TheoryStates}
\end{figure}
\review{\indent Since hybrid functional calculations are significantly more costly than GGAs (especially in case of plane wave calculations), an often used approach in the literature is to optimize the geometry at the PBE level, followed by a single SCF run at the HSE06 level for more a accurate electronic structure (indicated here as HSE@PBE).\cite{ekimov:2018DRM, VanWijkVanpouckeDEP:2025Carbon,thiering:2018prx, LofgrenR:NewJPhys2018} However, the PBE equilibrium volume is about $2$\% larger than what one would obtain for HSE06. Using a Rose--Vinet fit on a set of HSE@PBE energies, the HSE06 optimal equilibrium volumes are determined to be $97.91$\%($97.89$\%) and $97.82$\%($97.89$\%) of the PBE equilibrium volume for SnV$^0$ and SnV$^-$ using a $512$($1000$) atom supercell. The atomic structure at fixed volume is then optimized at the HSE06 level (indicated as HSE@HSE).\\} 
\indent Atomic charges are calculated using the Hirshfeld-I atoms-in-molecules partitioning scheme\cite{VanpouckeDannyEP:2013aJComputChem,VanpouckeDannyEP:2013bJComputChem,HIVE_REF}, using a convergence criterion of $1.0\times10^{-5}$ electron. Charges are integrated using an atom-centered Lebedev--Laikov grid of $1202$ grid-points per shell\cite{LebedevVI_grid:1999DokladyMath}.

\subsection{Color Center Model}\label{sec:model}
The Tin-split-Vacancy (SnV) color center is modeled by replacing a pair of C atoms by a single Sn atom located at the former bond center of the removed C atoms. After structure optimization, a six-coordinated Sn with $D_{3d}$ symmetry is obtained, as shown in Fig.~\ref{fig:TheoryStates}a. A conventional supercell of $512$ and $1000$ atoms is used, corresponding to a color center concentration of $0.195$ and $0.100$\%, respectively. The neutral, SnV$^0$, and negative charge state, SnV$^-$, are considered, and created by manually setting the number of electrons in the system. It should be noted that in both cases the color center draws in electrons from the host system resulting in an effectively negatively charged color center with $Q_{SnV^0}=-1.96$ and $Q_{SnV^-}=-2.54$ (\textit{cf.}, Table SI.I). This reflects a spatial charging of the color center region, which is not (necessarily) associated with the occupancy of certain electronic states. In this work, the SnV$^-$ thus refers to the latter, and not the spatial charging of the color center.\\
\indent To determine the color center character of the one-electron Kohn--Sham (KS) states, they are decomposed by projection onto atomic orbitals\cite{SchulerM:2018JPhysCondMat}. The resulting decomposition is then summed per atom, resulting in a single fractional contribution (per atom) to the KS one-electron state. In the current work, the Sn atom and its six neighboring C atoms are considered to constitute the color center (\textit{cf.}, Fig.~\ref{fig:TheoryStates}a). In previous work on the vibrational spectrum of point-defects in diamond, it was shown that beyond the first shell the host character is quickly regained\cite{VanpoukeDEP:ComputMaterSci2020}. As such, the sum-contribution of these seven atoms is used to provide a qualitative measure for the color center character of a KS state. This provides a simple way to identify the by-group-theory determined states of the color center, and distinguish them from host states.\review{Since the relevant states and excitations only involve the spin-minority channel (\textit{c.q.}, spin down in Fig.~\ref{fig:TheoryStates}), all discussed KS states and their tabulated values in SI will only refer to the states in the spin-minority channel, unless explicitly noted differently.}\\
\indent To investigate the impact of hydrostatic strain, a series of five supercells is created, containing the equilibrium volume of the electronic ground state (PBE-level) and volumes at $\pm1$ and $\pm2$\% of this equilibrium volume. Structures are optimized under the constraint of a fixed volume and a cubic lattice. The resulting energy-volume sets are fitted to the Rose--Vinet equation of state\cite{VinetP:PhysRevB1987_RoseVinetEOS, HIVE_REF}, allowing the determination of the pressure at each volume, and thus ZPL pressure coefficient.
\subsection{Zero-Phonon Line Approximations}\label{sec:ZPLmodel}
\begin{figure}[!tb]
    \centering
    \includegraphics[width=1\linewidth]{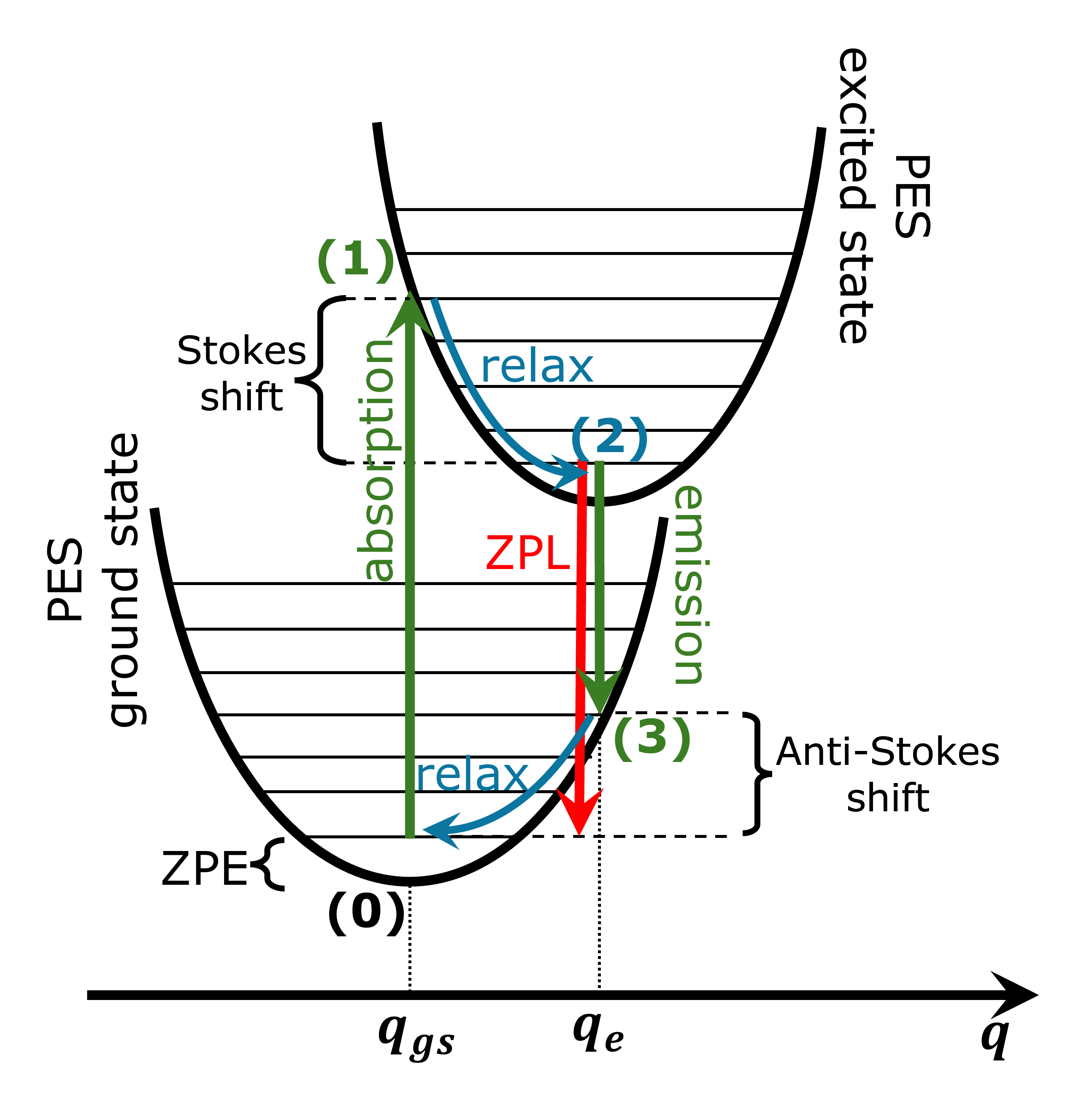}
    \caption{Schematic representation of the Franck--Codon approximation of the ZPL of a color center. }
    \label{fig:FranckCodon}
\end{figure}
Although this work focuses on the position of the SnV ZPL and its shift due to hydrostatic strain, we also want to draw some attention to the computational cost of different approaches and the resulting carbon footprint in relation to the actual accuracy of the calculations. For this reason, two popular functionals of the GGA and hybrid families (PBE and HSE06) are used in combination with two approximations for the calculation of the ZPL ($\Delta$KS and the Franck--Codon approximation (FC)).\\
\indent The $\Delta$KS approximation of the ZPL considers the ZPL as the energy difference between the KS states involved in the emission/absorption\cite{GossJP:2005PRB, Mavrin:2018JExpTheorPhys}. In case of the group-IV-vacancy color centers these are the $e_u$ and $e_g$ states, both as obtained in the  ground state of the system. As such, a single ground state structure optimization is all that is required.\\
\indent Within the FC approximation, the ZPL is determined as the total energy difference between the ground state configuration (0), and the total energy of a structurally relaxed excited state configuration (2) in Fig.~\ref{fig:FranckCodon}. Assuming single determinant wavefunctions with ground state electronic configuration $e_u^4e_g^2$ ($e_u^4e_g^3$) and excited state electronic configuration $e_u^3e_g^3$ ($e_u^3e_g^4$) for SnV$^0$ (SnV$^-$), the delta-self-consistent-field ($\Delta$SCF) approximation is used\cite{thiering:2018prx}. This means one $e_u$ electron (\textit{i.e.}, in an occupied KS state) is promoted to an empty $e_g$ state, and the electronic structure and geometry are optimized under this constraint to reflect the excited state system, effectively doubling the required computational cost compared to the $\Delta$KS approximation. Since the FC approximation requires total energies, these should be well converged quantities at the supercell level with regard to Brillouin-zone sampling and kinetic energy cut-off. 

\section{Results and Discussion}
\review{The results and their discussion are organized according to increasing complexity. First, in Sec.~\ref{sec:eupos} the identification of the relevant KS states of the SnV center is discussed, highlighting practical challenges. Then in Sec.~\ref{sec:ZPLposAll} the ZPL position is determined using two intrinsically different approximations. Their limitations and potential solutions are presented. The ZPL pressure coefficient under hydrostatic pressure as obtained from the two approximations are presented and compared in Sec.~\ref{sec:ZPLprescoef}. Finally, in Sec.~\ref{sec:compcost} the calculation of the ZPL position and shift is considered from the perspective of computational cost, as proxy for carbon footprint. Each section will be introduced with a brief presentation of the resulting conclusions, providing the experimental reader a birds eye perspective. The theoretical reader is invited to dive deeper into the technical discussion which follows.}

\subsection{The position of the $e_u$ KS states.\label{sec:eupos}}
\review{Although the group-theoretical picture of the electronic states of a group-IV-vacancy color center is well established\cite{GossJP:2005PRB,thiering:2018prx,VanWijkVanpouckeDEP:2025Carbon}, it does not provide clear information on the positioning of the states compared to the valence band maximum (VBM) and other host states (\textit{cf.}, Fig.~\ref{fig:TheoryStates}b). Within the context of periodic color center models, the position relative to the VBM of the $e_u$ KS state is shown to depend on the color center concentration, charge state, as well as excitation of a near VBM electron. As a result, the $e_u$ KS state of the SnV center is found to exchange order with the VBM-1 and even VBM KS state, making it essential to clearly identify this state correctly when investigating the ZPL.\\}
\indent Fortunately, by projecting the KS states onto atomic orbitals, the color center character of the KS states can be determined. For the SnV$^0$ color center, the $e_u$ states are located just below the VBM and a two-fold degenerate host state (indicated as VBM-1 in Fig.~\ref{fig:TheoryStates}c). \review{Similar as noted in previous work on the GeV$^0$\cite{VanWijkVanpouckeDEP:2025Carbon}, the relative position of the $e_u$ state depends on the color center concentration, with its position moving upward toward the VBM with decreasing SnV concentration.} For both the SnV$^0$ and SnV$^-$ ground state, the $e_u$ KS state is located below the VBM-1 states in case of the $512$ atom supercell. The VBM and VBM-1 KS states are separated by about $60$ meV at both the PBE and HSE06 level of theory (\textit{cf.}, Fig.~\ref{fig:TheoryStates}c), while for the $1000$ atom supercell this separation is halved. In contrast, the position of the $e_u$ KS state depends strongly on the color center charge state. For the SnV$^0$ and SnV$^-$, it is located $0.400$ eV and $0.078$ eV below the VBM, respectively, in case of the $512$ atom supercell at the PBE level (\textit{cf.}, Table SI.II). The $e_u$ states move upward with increasing supercell size, and for the $1000$ atom supercell are located $0.350$ eV and $0.009$ eV below the VBM (\textit{cf.}, Table SI.III). The latter now being located above the VBM-1 states. \review{This picture is further complicated when different occupancies near the VBM are considered to represent excited state systems. Using the PBE functional and the $512$ atom supercell, a nearly exhaustive set of excitation configurations is considered. The KS character and position is traced through a series of 3 calculations relevant for the Franck--Codon approximation, in addition to the ground state calculation (0) (\textit{cf.}, Fig.~\ref{fig:FranckCodon}): (1) vertical excitation (absorption) retaining the ground state atomic geometry, (2) relaxation of the atomic geometry after vertical excitation, and (3) vertical emission retaining the excited state geometry.\\
\indent At stage (1) and (3) of the Franck--Codon approximation, the atomic structure is not optimized for its respective electron configuration leading to the largest forces on an atom in the system of $0.2$--$1.4$ eV/\AA, as well as a potential significant reordering of the KS states which varies with the specific configuration (\textit{cf.}, Tables~SI.III and SI.IV). Upon relaxation (stage (2)), the maximum forces on a single atom reduce to values well below $1.0$ meV/\AA, indicative of a fully relaxed structure. The KS order, on the other hand, is consistent over the different configurations, taking the order of the ground state configuration for SnV$^0$, while shifting the $e_u$ states above the VBM for SnV$^-$. The reordering of the KS states originates mainly from the shift in the $e_u$ state position. While the host material KS states (VBM and VBM-1) show variations of about $10$ meV, the $e_u$ states move upward by $0.1$--$0.4$ eV, with the smallest shift occurring when an $e_u$ electron is excited (\textit{cf.}, Tables~SI.III and SI.IV).\\}
\indent A qualitatively similar picture is obtained for the HSE06 functional, though the $e_u$ states are already position above the VBM-1 states for the SnV$^-$ charge state using the $512$ atom supercell. The HSE06 functional furthermore gives rise to the added complication that a non-degenerate occupancy of the $e_g$ states also leads to a lifting of the degeneracy of the $e_u$ states (\textit{cf.}, Table SI.II and SI.III), deviating from the theoretical expectation.\\
\indent \review{Furthermore, also note that the position of the states also differs between the minority and majority-electron channels (for both PBE and HSE06 functionals), leading to a further deviation from the simplified theoretical picture, where both channels are represented as iso-energetic. This highlights an important challenge when dealing with this type of defect: the correct identification of the $e_u$ state.}\\

\begin{table}[!tb]
    \caption{The position of the ZPL in the $\Delta$KS model of the SnV center using the PBE and HSE06 functionals and supercells of $512$ ($0.195$\%) and $1000$ ($0.100$\%) atoms (color center concentration). ZPL position is given in nm units, and the experimentally observed peak position is considered to be at $619$ nm.}\label{table:ZPLeq_DKS}
    \begin{ruledtabular}
    \begin{tabular}{ll|rrr}
         & functional & $0.195$\% & $0.100$\% & dilute limit \\
         \hline
         SnV$^0$ & PBE     & $667$ & $701$ & $737$  \\
         SnV$^0$ & HSE@PBE & $497$ & $506$ & $515$  \\
         SnV$^0$ & HSE@HSE & $486$ & $495$ & $505$  \\
         SnV$^-$ & PBE     & $643$ & $650$ & $659$  \\
         SnV$^-$ & HSE@PBE & $491$ & $495$ & $499$  \\
         SnV$^-$ & HSE@HSE & $479$ & $483$ & $486$  \\
    \end{tabular}
    \end{ruledtabular}
\end{table}
\subsection{The unstrained SnV ZPL position.}\label{sec:ZPLposAll}
\review{Calculation of the absolute ZPL positions is a central task in predictive assignment of color center structures. The $\Delta$KS approximation which relies on the KS orbital positions shows a clear dependence on the supercell size, with the PBE functional providing good predictions in the dilute limit. Interestingly, we find that the self consistent HSE@HSE approach worsens the HSE@PBE predictions as a consequence of the reduced supercell size.\\
\indent The FC approximation, which relies on total energies, shows that a Brillouin-zone sampling beyond $\Gamma$-point only is essential, even for large supercells. It, however, also shows a consistent convergence to a unique ZPL, equal to the dilute limit of $\Gamma$-only calculations. The results at the PBE level are slightly redshifted. Although exact ZPL values may be hard to obtain from DFT, we observe that relative positions may be better represented, allowing us to propose that the $663$ nm ZPL in SnV samples might originate from SnV$^0$.\\}
\subsubsection{ZPL in the $\Delta$KS approximation}
Using the $\Delta$KS approximation, the ZPL position shows a clear dependence on the color center concentration. From the GeV it is known there exists a linear relation between the ZPL, expressed in eV, and concentration, expressed as percentage, allowing for the extrapolation to the dilute limit (\textit{c.q.}, $0$\%). At the PBE level a $512$ atom supercell provides a result of $643$ nm, close to the experimental value of $619$ nm. However, the true PBE result at experimental concentrations would be further redshifted with a value of $659$ nm, although still in quite good agreement. This highlights the risk of fortuitous agreement due to the selected system size.\\ 
\review{\indent Using the hybrid HSE06 functional, the HSE@PBE and HSE@HSE results present the same qualitative picture of the KS state positions. However, for the HSE@HSE results the absolute positions of the $e_u$ and $e_g$ states shifts upward $0.3$--$0.4$ eV for the $512$ atom supercell compared to the HSE@PBE case. This shift is slightly larger in case of the SnV$^-$. Because the shift in absolute position is not uniform for all KS states, the ZPL energy effectively increases by about $60$ meV, blueshifting away from the experimental expectation. Extrapolation to the dilute limit results in a partially compensating redshift, back to the high concentration HSE@PBE ZPL (\textit{cf.}, Table~\ref{table:ZPLeq_DKS}).\\
\indent Comparing the ZPL values of PBE and HSE06, a roughly constant shift of $0.65$ eV and $0.60$ eV for the SnV$^0$ and SnV$^-$ system is observed (\textit{cf.}, Table~SI.IV), similar as for the GeV color center\cite{VanWijkVanpouckeDEP:2025Carbon}.} In contrast to the PBE functional, the HSE06 functional severely overestimates the ZPL energy, predicting a SnV$^-$ ZPL at $500$ nm (\textit{cf.}, Table~\ref{table:ZPLeq_DKS}). It is important to note that in the current work, the KS states with $e_u$ character are considered, rather than the VBM, which gives rise to a ZPL with higher energy than previously reported\cite{MaryJoyVanpoucke:ACSMaterSci2026}. Furthermore, the theoretical value of $2.09$ eV presented in the literature for a $512$ atom system using HSE06\cite{thiering:2018prx} corresponds to the $\Delta$KS result between the $e_u$ and filled $e_g$ state for SnV$^-$ with non-degenerate configuration (\textit{cf.}, Table SI.II). However, the empty $e_g$ state is located $0.5$ eV higher, placing the associated ZPL significantly above the experimental value.\\
\review{\indent Thus the $\Delta$KS approximation predicts ZPL positions with a strong supercell size dependence, while the difference in ZPL between functionals corresponds to a constant shift using an energy scale.}

\subsubsection{ZPL in the Franck--Codon approximation}\label{sec:ZPL_FC}
A theoretically more accurate description of the ZPL can be obtained via the Franck--Codon approximation, using the $\Delta$SCF approximation, shown in Fig.~\ref{fig:FranckCodon}\cite{LofgrenR:NewJPhys2018,VindoletB:PRB2022}. In the current case, a single electron is (manually) promoted from a filled $e_u$ to an empty $e_g$ KS state, and the KS state occupancy is constrained during the SCF and structure optimizations. In practice, one needs to be rather careful as the occupancies are linked to KS state indices rather than KS states\cite{fn:KSindices}. As such, the expected character of the involved KS states should be validated at the end of the SCF calculation. As discussed in Sec.~\ref{sec:eupos}, selecting the appropriate KS state for excitation in the context of a $\Delta$SCF approach is not trivial. Investigation of the different excitation configurations shows that mainly the character of the KS state from which an electron is excited determines the absorption, emission, and ZPL energies, with clearly distinct values associated with the different KS state characters. In the case of SnV$^0$, excitation of an electron with $e_u$, VBM-1, or VBM character gives rise to a ZPL at $674.5$ ($693.8$), $725.4$ ($736.0$), and $758.0$ ($753.3$) nm, respectively, for a $512$ ($1000$) atom supercell (\textit{cf.}, Table SI.VII). In case of the SnV$^-$, the ZPL are located at $656.8$ ($664.2$), $583.6$ ($567.6$), and $607.4$ ($582.4$) nm, for excitation from an $e_u$, VBM-1, or VBM state, respectively (\textit{cf.}, Table SI.VIII).\\ 
\indent From these results, we observe a clear qualitative difference for excitations of different character electrons. The ZPL due to an $e_u$ electron excitation presents a blueshift for increasing color center concentration, while excitation from the VBM-1 and VBM KS states, presents a redshift for increasing color center concentration. Translated to the practical experimental context this means that the presence of other SnV colorcenters nearby (\textit{i.e.}, increasing the local density) should induce a blueshift of the ZPL. This has also been observed in theoretical previous work on the NV$^-$ and GeV$^0$ color centers\cite{LofgrenR:NewJPhys2018,VanWijkVanpouckeDEP:2025Carbon}. \review{In addition, it is also interesting to note that the ZPL due to the excitation of an electron in an $e_u$ KS state only differs $50$ meV between the SnV$^0$ and SnV$^-$. The Stokes and anti-Stokes shifts are symmetric as expected, and have a value of $30$--$40$ meV for an $e_u$ electron excitation. In contrast, excitations from the VBM and VBM-1 states differ $0.4$ eV between the charge states (\textit{cf.}, Tables~SI.VII and SI.VIII). This shows a clear difference in the qualitative picture exists depending on the charge state of the color center and the importance of correctly identifying the character of the KS state from which an electron is excited in a constrained $\Delta$SCF approximation.\\} 
\indent Using the HSE06 functional, qualitatively similar results are obtained, be it at a much higher computational cost. In addition, there is the practical complication that the two-fold degeneracy of the $e_u$, VBM-1, and $e_g$ states is lifted for each state, in case a non degenerate occupancy of any of these states is present, leading to different total energies\cite{fn:HSEHSE}. This is in stark contrast to the PBE calculations where degeneracies are retained, and different equivalent configurations have the same total energies. As shown in Table \ref{table:ZPLeq_FC_nm}, the PBE results may even provide a better representation of the experimental values. \review{Similar as for the $\Delta$KS approximation, extrapolation to the dilute limit results in a redshift of the ZPL.\\} 
\begin{table}[!tb]
    \caption{The position of the ZPL in the Franck--Codon model of the SnV center using the PBE and HSE06 (HSE@HSE approach) functionals and supercells of $512$ ($0.195$\%) and $1000$ ($0.100$\%) atoms (color center concentration). ZPL position is given in nm units, and the experimentally observed peak position is considered to be at $619$ nm.}\label{table:ZPLeq_FC_nm}
    \begin{ruledtabular}
    \begin{tabular}{ll|rrr}
         & functional & $0.195$\% & $0.100$\% & dilute limit \\
         \hline
         SnV$^0$ & PBE   & $674.5$ & $693.8$ & $715.2$  \\
         SnV$^0$ & HSE06  & $672.9$ & - & -  \\
         SnV$^-$ & PBE   & $656.8$ & $664.2$ & $672.0$  \\
         SnV$^-$ & HSE06-deg  & $547.0 $ & - & -  \\
         SnV$^-$ & HSE06-nondeg  & $598.1 $ & - & -  \\
    \end{tabular}
    \end{ruledtabular}
\end{table}

\subsubsection{Impact of Brillouin-zone sampling on the Franck--Codon ZPL.}
\indent In the current state-of-the-art literature, the first Brillouin-zone of systems with a size of $512$ and $1000$ atoms are sampled using the $\Gamma$-point only. This is then based on the argument that the energies are sufficiently converged as they vary less than $\leq1$ meV/atom, which is a sound argument in the context of bulk structural stability. However, the ZPL based on the Franck--Codon approximation is calculated as a difference of total energies. This means that an error of $1$ meV/atom corresponds to a $0.5$--$1.0$ eV error for the total energies used. The effect can be expected to be somewhat alleviated by a cancellation of errors, but not entirely. To investigate the impact on the ZPL position, $2\times2\times2$ and $3\times3\times3$ $\Gamma$-centered k-point grids for the $512$ and $1000$ atom systems using PBE, are also considered. For simplicity, the atomic geometries of the $\Gamma$-only calculations are retained. The resulting ZPL positions calculated according to the FC approximation are presented in Tables \ref{table:ZPLeq_FCkpoint_nm} and SI.IX. Where the total energies show a convergence of about $0.5$ eV ($0.2$ eV) (\textit{i.e.},$\leq1$ meV/atom), the ZPL position is determined more accurately, though still showing an error of $50$--$100$ meV ($20$--$50$ meV), for the $512$ ($1000$) atom supercell. More interestingly, the ZPL positions appear to converge to the same values as found for the dilute limit extrapolation (\textit{cf.}, Table~\ref{table:ZPLeq_FC_nm}), irrespective of the supercell size. It should be noted however, that when using an extended k-point set, care should be taken that for each individual k-point the excitation happens from the same band rather than same band index, as the relatively flat color center states cross the diamond KS VBM and VBM-1 states. Furthermore, in contrast to the position of the diamond KS states at the $\Gamma$-point, the color center states shift by a few tens of meV when an extended k-point set is used. Fortunately, this behaviour very quickly converges once going beyond the $\Gamma$-only sampling of the first Brillouin-zone.\\ 
\review{\indent Considering the different scaling of the computational cost with system size and Brillouin-zone sampling, these results may indicate that higher quality results might be achievable at lower cost when combining smaller supercells with an extended k-point set. It would however require extra vigilance when defining the excited KS state for the $\Delta$SCF approximation. Further work in this direction will be considered in future research.\\}
\begin{table}[!tb]
    \caption{The position of the ZPL in the Franck--Codon model of the SnV center using the PBE functional and a $512$($0.195$\%) or $1000$($0.100$\%) atom (color center concentration) supercell as function of the k-point set. ZPL position is given in nm units, and the experimentally observed peak position is considered to be at $619$ nm.}\label{table:ZPLeq_FCkpoint_nm}
    \begin{ruledtabular}
    \begin{tabular}{ll|ccc}
          & size & $\Gamma$ & $2\times2\times2$ & $3\times3\times3$  \\
         \hline
         SnV$^0$ & $512$ & $674.5$ & $725.9$ & $713.2$  \\
         SnV$^-$ & $512$ & $656.8$ & $675.1$ & $673.7$  \\
         \hline
         SnV$^0$ & $1000$ & $693.8$ & $724.2$ & $716.1$  \\
         SnV$^-$ & $1000$ & $664.2$ & $670.5$ & $670.5$  \\  
    \end{tabular}
    \end{ruledtabular}
\end{table}
\subsubsection{Charge state dependence of the SnV ZPL in the Franck--Codon approximation.}
\indent Based on the current results, it is clear that predicting the exact experimental ZPL position with nm-precision may be beyond the scope of current DFT methods. Close agreement seems to be a consequence of a fortunate choice of supercell, rather than theoretically converged results. However, considering general trends (and better cancellation of errors) may provide valid insights into the relative position of the ZPL for different charge states. In case of the SnV, the ZPL of the neutral charge state, SnV$^0$, is expected to be redshifted by about $43$ nm compared to the negative charge state, SnV$^-$, which seems reasonable in light of known experimental results for other color centers. The value is five times smaller than the observed redshift for the SiV, with a ZPL at $946$ and $738$ nm for the SiV$^0$ and SiV$^-$, respectively\cite{Haesens:2011PhysRevB_SiV0vsSiVminZPL, Green:2017PhysRevLett_SiV0ZPL}. It is closer to $20$ nm redshift for the GeV, if it is correct to attribute the $626$ nm ZPL to the GeV$^0$\cite{KrivobokV:2020PRB}. In contrast, the NV$^0$ ZPL is blueshift compared to the NV$^-$ ZPL, with experimentally observed values of $575$ and $637$ nm, respectively\cite{greentree:2008MatToday_NV}. In comparison, Ekimov \textit{et al.}\cite{ekimov:2018DRM} calculated values indicating a blueshift of $8$ nm of the SnV$^0$ compared to the SnV$^-$. This stands in stark contrast with the experimental observation of additional ZPL lines at $647$ nm observed by Tchernij \textit{et al.}\cite{tchernij:2017acs} and Corte \textit{et al.}\cite{Corte:2021AdvPhotonRes_SnV0}. They showed these to come from SnV sites, and suggested to attribute this ZPL to the SnV$^0$, which would mean the latter to be redshifted by $27$ nm compared to the SnV$^-$, which would be in reasonable agreement with the results presented here. However, other work by L\"{u}hman \textit{et al.}\cite{Luhman:2020ACSPhoton_ChargeTuneSnV} seems to indicate that the $647$ nm ZPL is related to a different Sn-based color center, though of unknown atomic configuration, highlighting the need for accurate and predictive methods of determining the exact ZPL position. \review{In recent work, Mary Joy \textit{et al.}\cite{MaryJoyVanpoucke:ACSMaterSci2026} also reported emission lines at $647$ nm and $663$ nm. The latter presenting a redshift of $43$ nm compared to the SnV$^-$ ZPL, making it an excellent candidate SnV$^0$ ZPL, though further experiments will be needed.\\}

\subsection{SnV ZPL pressure coefficient under hydrostatic strain.}\label{sec:ZPLprescoef}
In recent years, and with a focus on practical technological applications, the interest in the impact of strain on the ZPL position has gained steady theoretical interest\cite{VanWijkVanpouckeDEP:2025Carbon, MaryJoyVanpoucke:ACSMaterSci2026, VindoletB:PRB2022,ekimov:2018DRM, Karapatzakis:2024PhysRevX_GSsplit}. Considering the computational cost of large scale DFT calculations, the pressure coefficients are compared for the $\Delta$KS and FC approximations. \review{Where the $\Delta$KS approximation shows functional and charge state dependence, the FC predicted pressure coefficient of about $1.4$ nm/GPa appears nearly independent of the functional, charge state and even supercell size. In contrast to this difference, both approximations give rise to the same qualitative picture of a ZPL blueshift under increasing pressure, with only very slightly higher values for the negative charge state.}\\
\indent Within the $\Delta$KS approximation, the ZPL position is calculated as the difference between the energy of the $e_u$ and $e_g$ KS states at five volumes (\textit{i.e.}, PBE equilibrium, $\pm1$\% and $\pm2$\%). In case of the HSE@PBE results, the degeneracy of the $e_u$ and $e_g$ KS states is lifted, and the highest occupied $e_u$ state and the lowest unoccupied $e_g$ states are used. The results presented in Table~\ref{table:ZPLshift_dKZ_nmGPa} and SI.X are the result of a linear fit, with $R^2>0.99$. Since the absolute positions of the KS states show a linear relationship with the pressure, they are also linearly extrapolated to the dilute limit providing an estimate of the pressure coefficients at experimentally relevant ppm and ppb regimes. Both PBE and HSE06 present a similar qualitative picture of a blueshift with increasing pressure. The pressure coefficients are almost identical for the two charge states, with the negative charge state showing a slightly higher pressure coefficient. Furthermore, as seen in previous work on the GeV color center\cite{VanWijkVanpouckeDEP:2025Carbon}, the HSE06 functional gives rise to a significantly smaller pressure coefficient. In contrast to the ZPL position, the pressure coefficient is already well converged using $\Gamma$-only calculations.\\  
\begin{table}[!tb]
    \caption{The ZPL pressure coefficients using the $\Delta$KS approximation using the PBE and HSE06 functionals and supercells of $512$ ($0.195$\%) and $1000$ ($0.100$\%) atoms (color center concentration). The pressure coefficient is in nm/GPa units.}\label{table:ZPLshift_dKZ_nmGPa}
    \begin{ruledtabular}
    \begin{tabular}{lll|rrr}
         & functional & k-points & $0.195$\% & $0.100$\% & dilute limit \\
         \hline
         SnV$^0$ & PBE & $\Gamma$           & $1.34$ & $1.27$ & $1.15$  \\
         SnV$^0$ & PBE & $2\times2\times2$  & $1.35$ & $1.26$ & $1.15$  \\
         SnV$^-$ & PBE & $\Gamma$           & $1.37$ & $1.35$ & $1.34$  \\
         SnV$^-$ & PBE & $2\times2\times2$  & $1.36$ & $1.34$ & $1.33$  \\
         SnV$^0$ & HSE@PBE & $\Gamma$       & $0.88$ & $0.85$ & $0.81$  \\
         SnV$^-$ & HSE@PBE & $\Gamma$       & $0.91$ & $0.90$ & $0.89$  \\
    \end{tabular}
    \end{ruledtabular}
\end{table}
\indent The pressure coefficients have also been calculated using the Franck--Codon approximation, shown in Table~\ref{table:ZPLshift_FC_nmGPa} and SI.XI. Several qualitative differences compared to the $\Delta$KS approximation stand out. Firstly, the pressure coefficients appear to be converged for the $512$ atom supercell. Second, and more interestingly, the HSE06 results, have the same magnitude as the PBE results. Similarly as for the $\Delta$KS approximation, the $\Gamma$-only calculations already present converged results. A third and final qualitative difference is found in the observation that both the SnV$^0$ and SnV$^-$ show the same pressure coefficient, which means that their ZPL lines will shift identically under hydrostatic pressure. Based on the results presented, a pressure coefficient of $1.4$ nm/GPa is expected for the SnV color center, which is larger than the value of $0.4$--$0.8$ nm/GPa reported for the GeV color center\cite{VanWijkVanpouckeDEP:2025Carbon,VindoletB:PRB2022,KrivobokV:2020PRB,ekimovE:2019jetp}.\\
\indent A small side-note to consider from the computational perspective is the role of numerical settings on absolute energies. Although it is easy to convert the units of the pressure coefficient between meV/GPa and nm/GPa, starting from calculated total energies or KS state energies, use of nm/GPa seems to provide more robust results. Comparison of the results reported in Tables~\ref{table:ZPLshift_dKZ_nmGPa} and SI.X or Tables~\ref{table:ZPLshift_FC_nmGPa} and SI.XI, highlight variations of $1$--$2$ meV/GPa are found, though corresponding to (nearly) identical pressure coefficient when expressed in nm/GPa. This leads to the recommendation to use nm/GPa as primary unit, reducing the dependence of the results on numerical settings, allowing for better comparison to experimental observations.\\ 
\begin{table}[!tb]
    \caption{The ZPL pressure coefficients using the Franck--Codon approximation using the PBE and HSE06 functionals and supercells of $512$ ($0.195$\%) and $1000$ ($0.100$\%) atoms (color center concentration). The pressure coefficient is in nm/GPa units.}\label{table:ZPLshift_FC_nmGPa}
    \begin{ruledtabular}
    \begin{tabular}{lll|rrr}
         & functional & k-points & $0.195$\% & $0.100$\% & dilute limit \\
         \hline
         SnV$^0$ & PBE & $\Gamma$           & $1.40$ & $1.39$ & $1.36$  \\
         SnV$^0$ & PBE & $2\times2\times2$  & $1.39$ & $1.34$ & $1.30$  \\
         SnV$^-$ & PBE & $\Gamma$           & $1.41$ & $1.39$ & $1.39$  \\
         SnV$^-$ & PBE & $2\times2\times2$  & $1.41$ & $1.39$ & $1.39$  \\
         SnV$^-$ & HSE@PBE &$\Gamma$        & $1.39$ & $1.37$ & $1.37$  \\
    \end{tabular}
    \end{ruledtabular}
\end{table}

\subsection{Computational cost}\label{sec:compcost}
\indent Although first principles calculations provide a very powerful tool for prediction of materials properties, they also come at a significant computational cost, and consequently carbon footprint. In recent years, some have called for a shift in mindset toward frugal computing, where a problem is redefined or reformulated as to obtain the same quality results but at a much lower cost\cite{Gutierrez:2025GreenChem_Frugal}. Although the current topic appears not easily amenable to such a mindset, one can still remain critical of the quality obtained by the standard state-of-the-art approaches \review{and try to strike a balance between achievable accuracy and cost by considering `good enough' results\cite{BosoniE:NatRevPhys2023}. Therefore, in the second half of this section some recommendations based on the current work are presented.\\}
\indent The computational cost values presented here should not be considered absolute truths as they depend on the software package used, the compilation, and hardware, as well as chosen specific numerical settings and the system and electron-configuration under study. Furthermore, several tricks of the trade were used to reduce the computational cost such as the use of pre-converged (during another part of the project) structures, multi-step SCF cycles with increasing accuracy for the hybrid functional, specially optimized code for $\Gamma$-only calculations, etc. As such, the presented values are only meant to provide some intuition on the computational cost for a study like the one presented in this manuscript. To simplify comparison, the values are scaled relative to the cheapest calculation considered.\\
\indent The high cost of hybrid functionals is well known, as they represent a higher rung on Perdew's Ladder\cite{PerdewJacobs:AIPConfProc2001}. In the current case, it is further increased due to the use of a plane wave basis-set, and a slower convergence than PBE. Calculation using an extended k-point set were not performed for HSE06, as the cost would have been unacceptably high. The use of larger supercells also leads to the expected increase in computational cost (roughly cubic scaling), while the cost of using an extended k-point set scales linearly with the number of irreducible k-points at the PBE level of theory. The computational costs presented in Table~\ref{table:compcost} range from $64$ CPUh up to $46200$ CPUh for a single calculation, with the total computational cost for the current work exceeding $1.7\times10^6$ CPUh (equivalent to $987$ kg CO$_2$\cite{fn:CO2}).\\
\begin{table}[!tb]
    \caption{Computational cost for various calculations performed using $512$ and $1000$ atom systems, relative to the cost of a single SCF cycle at the PBE level for the $512$ atom system ($\sim 64$CPUh).}\label{table:compcost}
    \begin{ruledtabular}
    \begin{tabular}{lc|rrrr}
         & k-points & \multicolumn{2}{c}{PBE} & \multicolumn{2}{c}{HSE06} \\
         &          & $512$ & $1000$ & $512$ & $1000$ \\ 
         \hline
         SCF cycle  & $\Gamma$           & $1.0$         &  $6.0$ & $34.9$  & $312.8$ \\
         relaxation & $\Gamma$           & $8.9$         & $58.7$ & $176.4$ & $721.9$ \\
         SCF cycle  & $2\times2\times2$  & $5.4$         & $44.6$ & - & - \\
         SCF cycle  & $3\times3\times3$  & $8.7$--$62.5$ $^{\dag}$ & $67.2$ & - & - \\
    \end{tabular}
    \end{ruledtabular}
    $^{\dag}$ Range due to variation in irreducible k-points.
\end{table}
\indent Within this work, two properties are considered: the ZPL position and the pressure coefficient of the ZPL. For the former, it is clear that at the PBE level, both the $\Delta$KS and Franck--Codon approximations converge to similar results in the dilute limit when using a $\Gamma$-only k-point set. More interestingly, the dilute limit is also reached through the use of an extended k-point set, indicating that converged results can be reached based on a single supercell size without a need for extrapolation. Furthermore, it may even indicate that the smaller $216$ atom supercell with extended k-point set could suffice, significantly reducing the computational cost while maintaining the same precision. However, in the latter case, it is important to assign the color center character to the correct KS states in each point of the Brillouin-zone, as the dispersion of the host states leads to band crossing, effectively reordering the KS states at different k-points.\\ 
\indent The pressure coefficient, in contrast, shows a very robust behavior when employing the Franck--Codon approximation. In this case, $\Gamma$-only calculations suffice to reach convergence. From the computational cost perspective, it is also clear that the low rung PBE functional provides the exact same results as obtained for the higher rung hybrid functional, when considering nm/GPa units. \review{This makes PBE computationally cost effective for pressure coefficient calculations using the FC approximation.\\}
\section{Conclusion}
The position of the zero-phonon line(ZPL) of the SnV color center was predicted based on two models reflecting the physical process of vertical electron (de-)excitation. The first model considers only the Kohn--Sham(KS) states in the system ground state between which the electron excitation is to occur ($\Delta$KS). Although, it requires only a single ground state configuration to be optimized, there is a strong dependence on the functional, and system size, requiring an extrapolation to the dilute limit as first principles calculations are performed at $\geq1000$ ppm. Furthermore, results depend strongly on the functional employed, with PBE slightly redshifting the ZPL position, and HSE06 significantly blueshifting it. The second model instead considers the emission process between a relaxed excited configuration and the relaxed ground state: the Franck--Codon approximation (FC). As two structure optimizations are required, the computational cost doubles compared to the $\Delta$KS approximation. Using $\Gamma$-only sampling of the first Brillouin-zone also results in the need for an extrapolation to the dilute limit. For the PBE functional, both models show a similar redshift of the ZPL, while HSE06 calculations are complicated by a lifting of the expected degeneracy of KS states, with results blueshifted to varying extend. As the FC model is based on total energies, it is important to have these well converged at the level of the system. As such, one should not consider convergence at the level of energy per atom, since this provides a rather too optimistic picture of reality. For systems of $512$ and $1000$ atoms, total energies using $\Gamma$-only sampling have an accuracy of $0.5$ and $0.2$ eV, respectively. In contrast, a $\Gamma$-centered $2\times2\times2$ k-point sampling improves convergence to a few tens of meV in the worst case. More interestingly, the ZPL position converges to the dilute limit, removing the need for extrapolation based on different system sizes. This leads to the recommendation of using smaller supercells combined with an extended k-point sampling. Within the FC approximation, the HSE06 functional presents strongly blueshifted ZPL positions, which furthermore depend heavily on the specific electronic configuration considered, in contrast to PBE. In conclusion, it appears that DFT may not currently provide us with the means needed to determine absolute ZPL positions for color centers in diamond, with examples of exact agreement rather being fortuitous artifacts of the selected supercell and functional. However, relative positioning of differently charged color centers may be more accurate, as additional cancellation of errors is present. In the case of the SnV$^0$ color center, the ZPL is expected to be redshifted by about $43$ nm compared to the SnV$^-$ ZPL, thus predicting an experimental ZPL around $663$ nm for the SnV$^0$.\\
\indent For the two approximations, it is important to realize that the effective position of the KS state is strongly dependent on the charge state of the system, the color center concentration, the functional used, the k-point set considered, as well as the occupancy configuration. These need to be checked and validated thoroughly, as unfortuitous selection of a diamond host-state will result in different trends and absolute values. The reordering of states due to assigned occupancy is especially problematic for the HSE06 functional.\\
\indent In contrast to the absolute position of the ZPL, it was shown that the pressure coefficient is rather robust. A $\Gamma$-only sampling of the first Brillouin-zone is sufficient for both models. There is, however, a qualitative difference between the $\Delta$KS and FC models. The results of the former show a functional dependence, which is not the case for the latter. Furthermore, for the FC model the pressure coefficient appears to be converged at the $512$ atom supercell, and independent of the functional. In case of the SnV color center, a pressure coefficient of $1.40$ nm/GPa blueshift is found under increasing pressure for both the SNV$^0$ and SnV$^-$ color center, indicating both ZPL to exhibit similar behavior.

\section*{Acknowledgment}
\indent The author acknowledges the valuable discussion with R. Mary Joy on experimentally observed ZPL lines in SnV samples. He also wishes to dedicate this work to his recently departed sister K. Vanpoucke, in honor of her analytical and critical mindset, promoting STEM to the next generation. The computational resources and services used in this work were provided by the VSC (Flemish Supercomputer Center), funded by the Research Foundation Flanders (FWO) and the Flemish Government--department WEWIS, grant 2025\_098.

\section*{Data Availability}
The data that support the findings of this study are available from the corresponding author upon reasonable request.


\begin{thebibliography}{64}%
\makeatletter
\providecommand \@ifxundefined [1]{%
 \@ifx{#1\undefined}
}%
\providecommand \@ifnum [1]{%
 \ifnum #1\expandafter \@firstoftwo
 \else \expandafter \@secondoftwo
 \fi
}%
\providecommand \@ifx [1]{%
 \ifx #1\expandafter \@firstoftwo
 \else \expandafter \@secondoftwo
 \fi
}%
\providecommand \natexlab [1]{#1}%
\providecommand \enquote  [1]{``#1''}%
\providecommand \bibnamefont  [1]{#1}%
\providecommand \bibfnamefont [1]{#1}%
\providecommand \citenamefont [1]{#1}%
\providecommand \href@noop [0]{\@secondoftwo}%
\providecommand \href [0]{\begingroup \@sanitize@url \@href}%
\providecommand \@href[1]{\@@startlink{#1}\@@href}%
\providecommand \@@href[1]{\endgroup#1\@@endlink}%
\providecommand \@sanitize@url [0]{\catcode `\\12\catcode `\$12\catcode `\&12\catcode `\#12\catcode `\^12\catcode `\_12\catcode `\%12\relax}%
\providecommand \@@startlink[1]{}%
\providecommand \@@endlink[0]{}%
\providecommand \url  [0]{\begingroup\@sanitize@url \@url }%
\providecommand \@url [1]{\endgroup\@href {#1}{\urlprefix }}%
\providecommand \urlprefix  [0]{URL }%
\providecommand \Eprint [0]{\href }%
\providecommand \doibase [0]{https://doi.org/}%
\providecommand \selectlanguage [0]{\@gobble}%
\providecommand \bibinfo  [0]{\@secondoftwo}%
\providecommand \bibfield  [0]{\@secondoftwo}%
\providecommand \translation [1]{[#1]}%
\providecommand \BibitemOpen [0]{}%
\providecommand \bibitemStop [0]{}%
\providecommand \bibitemNoStop [0]{.\EOS\space}%
\providecommand \EOS [0]{\spacefactor3000\relax}%
\providecommand \BibitemShut  [1]{\csname bibitem#1\endcsname}%
\let\auto@bib@innerbib\@empty
\bibitem [{\citenamefont {Iv{\'a}dy}\ \emph {et~al.}(2018)\citenamefont {Iv{\'a}dy}, \citenamefont {Abrikosov},\ and\ \citenamefont {Gali}}]{GaliA:2018npjCompMat}%
  \BibitemOpen
  \bibfield  {author} {\bibinfo {author} {\bibfnamefont {V.}~\bibnamefont {Iv{\'a}dy}}, \bibinfo {author} {\bibfnamefont {I.~A.}\ \bibnamefont {Abrikosov}},\ and\ \bibinfo {author} {\bibfnamefont {A.}~\bibnamefont {Gali}},\ }\bibfield  {title} {\bibinfo {title} {First principles calculation of spin-related quantities for point defect qubit research},\ }\href {https://doi.org/10.1038/s41524-018-0132-5} {\bibfield  {journal} {\bibinfo  {journal} {npj Comput. Mater.}\ }\textbf {\bibinfo {volume} {4}},\ \bibinfo {pages} {76} (\bibinfo {year} {2018})}\BibitemShut {NoStop}%
\bibitem [{\citenamefont {Ruf}\ \emph {et~al.}(2021)\citenamefont {Ruf}, \citenamefont {Wan}, \citenamefont {Choi}, \citenamefont {Englund},\ and\ \citenamefont {Hanson}}]{RufM:2021JAppPhys}%
  \BibitemOpen
  \bibfield  {author} {\bibinfo {author} {\bibfnamefont {M.}~\bibnamefont {Ruf}}, \bibinfo {author} {\bibfnamefont {N.~H.}\ \bibnamefont {Wan}}, \bibinfo {author} {\bibfnamefont {H.}~\bibnamefont {Choi}}, \bibinfo {author} {\bibfnamefont {D.}~\bibnamefont {Englund}},\ and\ \bibinfo {author} {\bibfnamefont {R.}~\bibnamefont {Hanson}},\ }\bibfield  {title} {\bibinfo {title} {Quantum networks based on color centers in diamond},\ }\href {https://doi.org/10.1063/5.0056534} {\bibfield  {journal} {\bibinfo  {journal} {J. Appl. Phys.}\ }\textbf {\bibinfo {volume} {130}},\ \bibinfo {pages} {070901} (\bibinfo {year} {2021})}\BibitemShut {NoStop}%
\bibitem [{\citenamefont {Mukherjee}\ \emph {et~al.}(2023)\citenamefont {Mukherjee}, \citenamefont {Zhang}, \citenamefont {Oblinsky}, \citenamefont {de~Vries}, \citenamefont {Johnson}, \citenamefont {Gibson}, \citenamefont {Mayes}, \citenamefont {Edmonds}, \citenamefont {Palmer}, \citenamefont {Markham}, \citenamefont {Gali}, \citenamefont {Thiering}, \citenamefont {Dalis}, \citenamefont {Dumm}, \citenamefont {Scholes}, \citenamefont {Stacey}, \citenamefont {Reineck},\ and\ \citenamefont {de~Leon}}]{Mukherjee:2023NanoLett}%
  \BibitemOpen
  \bibfield  {author} {\bibinfo {author} {\bibfnamefont {S.}~\bibnamefont {Mukherjee}}, \bibinfo {author} {\bibfnamefont {Z.-H.}\ \bibnamefont {Zhang}}, \bibinfo {author} {\bibfnamefont {D.~G.}\ \bibnamefont {Oblinsky}}, \bibinfo {author} {\bibfnamefont {M.~O.}\ \bibnamefont {de~Vries}}, \bibinfo {author} {\bibfnamefont {B.~C.}\ \bibnamefont {Johnson}}, \bibinfo {author} {\bibfnamefont {B.~C.}\ \bibnamefont {Gibson}}, \bibinfo {author} {\bibfnamefont {E.~L.~H.}\ \bibnamefont {Mayes}}, \bibinfo {author} {\bibfnamefont {A.~M.}\ \bibnamefont {Edmonds}}, \bibinfo {author} {\bibfnamefont {N.}~\bibnamefont {Palmer}}, \bibinfo {author} {\bibfnamefont {M.~L.}\ \bibnamefont {Markham}}, \bibinfo {author} {\bibfnamefont {A.}~\bibnamefont {Gali}}, \bibinfo {author} {\bibfnamefont {G.}~\bibnamefont {Thiering}}, \bibinfo {author} {\bibfnamefont {A.}~\bibnamefont {Dalis}}, \bibinfo {author} {\bibfnamefont {T.}~\bibnamefont {Dumm}}, \bibinfo {author} {\bibfnamefont {G.~D.}\ \bibnamefont {Scholes}}, \bibinfo {author}
  {\bibfnamefont {A.}~\bibnamefont {Stacey}}, \bibinfo {author} {\bibfnamefont {P.}~\bibnamefont {Reineck}},\ and\ \bibinfo {author} {\bibfnamefont {N.~P.}\ \bibnamefont {de~Leon}},\ }\bibfield  {title} {\bibinfo {title} {{A Telecom O-Band Emitter in Diamond}},\ }\href {https://doi.org/10.1021/acs.nanolett.2c04608} {\bibfield  {journal} {\bibinfo  {journal} {Nano Lett.}\ }\textbf {\bibinfo {volume} {23}},\ \bibinfo {pages} {2557} (\bibinfo {year} {2023})}\BibitemShut {NoStop}%
\bibitem [{\citenamefont {Filippatos}\ \emph {et~al.}(2025)\citenamefont {Filippatos}, \citenamefont {Chroneos},\ and\ \citenamefont {Kelaidis}}]{Filippatos:2025JApplPhys}%
  \BibitemOpen
  \bibfield  {author} {\bibinfo {author} {\bibfnamefont {P.-P.}\ \bibnamefont {Filippatos}}, \bibinfo {author} {\bibfnamefont {A.}~\bibnamefont {Chroneos}},\ and\ \bibinfo {author} {\bibfnamefont {N.}~\bibnamefont {Kelaidis}},\ }\bibfield  {title} {\bibinfo {title} {A first-principles investigation of halogen doped diamond and its application to quantum technologies},\ }\href {https://doi.org/10.1063/5.0279139} {\bibfield  {journal} {\bibinfo  {journal} {J. Appl. Phys.}\ }\textbf {\bibinfo {volume} {138}},\ \bibinfo {pages} {094401} (\bibinfo {year} {2025})}\BibitemShut {NoStop}%
\bibitem [{\citenamefont {Damle}\ \emph {et~al.}(2020)\citenamefont {Damle}, \citenamefont {Wu}, \citenamefont {Luca}, \citenamefont {{n}}, \citenamefont {Norouzi}, \citenamefont {Morita}, \citenamefont {de~Vries}, \citenamefont {Kaper}, \citenamefont {Zuhorn}, \citenamefont {Eisel}, \citenamefont {Vanpoucke}, \citenamefont {Rudolf},\ and\ \citenamefont {Schirhagl}}]{DamleVVanpouckeDEP:2020bCarbon}%
  \BibitemOpen
  \bibfield  {author} {\bibinfo {author} {\bibfnamefont {V.}~\bibnamefont {Damle}}, \bibinfo {author} {\bibfnamefont {K.}~\bibnamefont {Wu}}, \bibinfo {author} {\bibfnamefont {O.~D.}\ \bibnamefont {Luca}}, \bibinfo {author} {\bibfnamefont {N.~O.-C.}\ \bibnamefont {{n}}}, \bibinfo {author} {\bibfnamefont {N.}~\bibnamefont {Norouzi}}, \bibinfo {author} {\bibfnamefont {A.}~\bibnamefont {Morita}}, \bibinfo {author} {\bibfnamefont {J.}~\bibnamefont {de~Vries}}, \bibinfo {author} {\bibfnamefont {H.}~\bibnamefont {Kaper}}, \bibinfo {author} {\bibfnamefont {I.~S.}\ \bibnamefont {Zuhorn}}, \bibinfo {author} {\bibfnamefont {U.}~\bibnamefont {Eisel}}, \bibinfo {author} {\bibfnamefont {D.~E.~P.}\ \bibnamefont {Vanpoucke}}, \bibinfo {author} {\bibfnamefont {P.}~\bibnamefont {Rudolf}},\ and\ \bibinfo {author} {\bibfnamefont {R.}~\bibnamefont {Schirhagl}},\ }\bibfield  {title} {\bibinfo {title} {Influence of diamond crystal orientation on the interaction with biological matter},\ }\href
  {https://doi.org/10.1016/j.carbon.2020.01.115} {\bibfield  {journal} {\bibinfo  {journal} {Carbon}\ }\textbf {\bibinfo {volume} {162}},\ \bibinfo {pages} {1} (\bibinfo {year} {2020})}\BibitemShut {NoStop}%
\bibitem [{\citenamefont {Mzyk}\ \emph {et~al.}(2022)\citenamefont {Mzyk}, \citenamefont {Ong}, \citenamefont {Ortiz~Moreno}, \citenamefont {Padamati}, \citenamefont {Zhang}, \citenamefont {Reyes-San-Martin},\ and\ \citenamefont {Schirhagl}}]{Schirhagl:2022AnalChem_revBiochem}%
  \BibitemOpen
  \bibfield  {author} {\bibinfo {author} {\bibfnamefont {A.}~\bibnamefont {Mzyk}}, \bibinfo {author} {\bibfnamefont {Y.}~\bibnamefont {Ong}}, \bibinfo {author} {\bibfnamefont {A.~R.}\ \bibnamefont {Ortiz~Moreno}}, \bibinfo {author} {\bibfnamefont {S.~K.}\ \bibnamefont {Padamati}}, \bibinfo {author} {\bibfnamefont {Y.}~\bibnamefont {Zhang}}, \bibinfo {author} {\bibfnamefont {C.~A.}\ \bibnamefont {Reyes-San-Martin}},\ and\ \bibinfo {author} {\bibfnamefont {R.}~\bibnamefont {Schirhagl}},\ }\bibfield  {title} {\bibinfo {title} {{Diamond Color Centers in Diamonds for Chemical and Biochemical Analysis and Visualization}},\ }\href {https://doi.org/10.1021/acs.analchem.1c04536} {\bibfield  {journal} {\bibinfo  {journal} {Anal. Chem.}\ }\textbf {\bibinfo {volume} {94}},\ \bibinfo {pages} {225} (\bibinfo {year} {2022})}\BibitemShut {NoStop}%
\bibitem [{\citenamefont {Neumann}\ \emph {et~al.}(2013)\citenamefont {Neumann}, \citenamefont {Jakobi}, \citenamefont {Dolde}, \citenamefont {Burk}, \citenamefont {Reuter}, \citenamefont {Waldherr}, \citenamefont {Honert}, \citenamefont {Wolf}, \citenamefont {Brunner}, \citenamefont {Shim}, \citenamefont {Suter}, \citenamefont {Sumiya}, \citenamefont {Isoya},\ and\ \citenamefont {Wrachtrup}}]{Neumann:2013nanoLett_DiamTermo}%
  \BibitemOpen
  \bibfield  {author} {\bibinfo {author} {\bibfnamefont {P.}~\bibnamefont {Neumann}}, \bibinfo {author} {\bibfnamefont {I.}~\bibnamefont {Jakobi}}, \bibinfo {author} {\bibfnamefont {F.}~\bibnamefont {Dolde}}, \bibinfo {author} {\bibfnamefont {C.}~\bibnamefont {Burk}}, \bibinfo {author} {\bibfnamefont {R.}~\bibnamefont {Reuter}}, \bibinfo {author} {\bibfnamefont {G.}~\bibnamefont {Waldherr}}, \bibinfo {author} {\bibfnamefont {J.}~\bibnamefont {Honert}}, \bibinfo {author} {\bibfnamefont {T.}~\bibnamefont {Wolf}}, \bibinfo {author} {\bibfnamefont {A.}~\bibnamefont {Brunner}}, \bibinfo {author} {\bibfnamefont {J.~H.}\ \bibnamefont {Shim}}, \bibinfo {author} {\bibfnamefont {D.}~\bibnamefont {Suter}}, \bibinfo {author} {\bibfnamefont {H.}~\bibnamefont {Sumiya}}, \bibinfo {author} {\bibfnamefont {J.}~\bibnamefont {Isoya}},\ and\ \bibinfo {author} {\bibfnamefont {J.}~\bibnamefont {Wrachtrup}},\ }\bibfield  {title} {\bibinfo {title} {{High-Precision Nanoscale Temperature Sensing Using Single Defects in Diamond}},\
  }\href {https://doi.org/10.1021/nl401216y} {\bibfield  {journal} {\bibinfo  {journal} {Nano Lett.}\ }\textbf {\bibinfo {volume} {13}},\ \bibinfo {pages} {2738} (\bibinfo {year} {2013})}\BibitemShut {NoStop}%
\bibitem [{\citenamefont {{van Wijk}}\ \emph {et~al.}(2025)\citenamefont {{van Wijk}}, \citenamefont {Melan}, \citenamefont {Joy}, \citenamefont {Guillaume}, \citenamefont {Pobedinskas}, \citenamefont {Haenen},\ and\ \citenamefont {Vanpoucke}}]{VanWijkVanpouckeDEP:2025Carbon}%
  \BibitemOpen
  \bibfield  {author} {\bibinfo {author} {\bibfnamefont {T.~G.}\ \bibnamefont {{van Wijk}}}, \bibinfo {author} {\bibfnamefont {E.~A.}\ \bibnamefont {Melan}}, \bibinfo {author} {\bibfnamefont {R.~M.}\ \bibnamefont {Joy}}, \bibinfo {author} {\bibfnamefont {E.~Y.}\ \bibnamefont {Guillaume}}, \bibinfo {author} {\bibfnamefont {P.}~\bibnamefont {Pobedinskas}}, \bibinfo {author} {\bibfnamefont {K.}~\bibnamefont {Haenen}},\ and\ \bibinfo {author} {\bibfnamefont {D.~E.~P.}\ \bibnamefont {Vanpoucke}},\ }\bibfield  {title} {\bibinfo {title} {The impact of strain on the {GeV}-color center in diamond},\ }\href {https://doi.org/10.1016/j.carbon.2024.119928} {\bibfield  {journal} {\bibinfo  {journal} {Carbon}\ }\textbf {\bibinfo {volume} {234}},\ \bibinfo {pages} {119928} (\bibinfo {year} {2025})}\BibitemShut {NoStop}%
\bibitem [{\citenamefont {Razgulov}\ \emph {et~al.}(2025)\citenamefont {Razgulov}, \citenamefont {Lyapin}, \citenamefont {Novikov},\ and\ \citenamefont {Ekimov}}]{Ekimov:2025PhysRevB_G4strainZPL}%
  \BibitemOpen
  \bibfield  {author} {\bibinfo {author} {\bibfnamefont {A.~A.}\ \bibnamefont {Razgulov}}, \bibinfo {author} {\bibfnamefont {S.~G.}\ \bibnamefont {Lyapin}}, \bibinfo {author} {\bibfnamefont {A.~P.}\ \bibnamefont {Novikov}},\ and\ \bibinfo {author} {\bibfnamefont {E.~A.}\ \bibnamefont {Ekimov}},\ }\bibfield  {title} {\bibinfo {title} {{Pressure effect on low-temperature photoluminescence of the SiV, GeV, and SnV color centers in high-pressure high-temperature diamonds}},\ }\href {https://doi.org/10.1103/PhysRevB.111.024114} {\bibfield  {journal} {\bibinfo  {journal} {Phys. Rev. B}\ }\textbf {\bibinfo {volume} {111}},\ \bibinfo {pages} {024114} (\bibinfo {year} {2025})}\BibitemShut {NoStop}%
\bibitem [{\citenamefont {Zaitsev}(2001)}]{Zaitsev:2001ColorCenters}%
  \BibitemOpen
  \bibfield  {author} {\bibinfo {author} {\bibfnamefont {A.~M.}\ \bibnamefont {Zaitsev}},\ }\href {https://doi.org/10.1007/978-3-662-04548-0} {\emph {\bibinfo {title} {{Optical Properties of Diamond}}}}\ (\bibinfo  {publisher} {Springer},\ \bibinfo {address} {Berlin},\ \bibinfo {year} {2001})\BibitemShut {NoStop}%
\bibitem [{\citenamefont {Guillaume}\ \emph {et~al.}(2025)\citenamefont {Guillaume}, \citenamefont {Melan},\ and\ \citenamefont {Vanpoucke}}]{VanpouckeDEP:2025BookChapColor}%
  \BibitemOpen
  \bibfield  {author} {\bibinfo {author} {\bibfnamefont {E.~Y.}\ \bibnamefont {Guillaume}}, \bibinfo {author} {\bibfnamefont {E.~A.}\ \bibnamefont {Melan}},\ and\ \bibinfo {author} {\bibfnamefont {D.~E.~P.}\ \bibnamefont {Vanpoucke}},\ }\bibfield  {title} {\bibinfo {title} {Ab initio simulations of color centers in diamond},\ }in\ \href@noop {} {\emph {\bibinfo {booktitle} {Nanophotonics with Diamond and Silicon Carbide for Quantum Technologies}}},\ \bibinfo {editor} {edited by\ \bibinfo {editor} {\bibfnamefont {M.}~\bibnamefont {Agio}}\ and\ \bibinfo {editor} {\bibfnamefont {S.}~\bibnamefont {Castelletto}}}\ (\bibinfo  {publisher} {Elsevier},\ \bibinfo {address} {Amsterdam},\ \bibinfo {year} {2025})\ pp.\ \bibinfo {pages} {77--99}\BibitemShut {NoStop}%
\bibitem [{\citenamefont {Goss}\ \emph {et~al.}(1996)\citenamefont {Goss}, \citenamefont {Jones}, \citenamefont {Breuer}, \citenamefont {Briddon},\ and\ \citenamefont {\"Oberg}}]{GossJP:1996prl}%
  \BibitemOpen
  \bibfield  {author} {\bibinfo {author} {\bibfnamefont {J.~P.}\ \bibnamefont {Goss}}, \bibinfo {author} {\bibfnamefont {R.}~\bibnamefont {Jones}}, \bibinfo {author} {\bibfnamefont {S.~J.}\ \bibnamefont {Breuer}}, \bibinfo {author} {\bibfnamefont {P.~R.}\ \bibnamefont {Briddon}},\ and\ \bibinfo {author} {\bibfnamefont {S.}~\bibnamefont {\"Oberg}},\ }\bibfield  {title} {\bibinfo {title} {{The Twelve-Line 1.682 eV Luminescence Center in Diamond and the Vacancy-Silicon Complex}},\ }\href {https://doi.org/10.1103/PhysRevLett.77.3041} {\bibfield  {journal} {\bibinfo  {journal} {Phys. Rev. Lett.}\ }\textbf {\bibinfo {volume} {77}},\ \bibinfo {pages} {3041} (\bibinfo {year} {1996})}\BibitemShut {NoStop}%
\bibitem [{\citenamefont {Thiering}\ and\ \citenamefont {Gali}(2021)}]{GaliThiering:2021PhysRevRes_NiV}%
  \BibitemOpen
  \bibfield  {author} {\bibinfo {author} {\bibfnamefont {G.}~\bibnamefont {Thiering}}\ and\ \bibinfo {author} {\bibfnamefont {A.}~\bibnamefont {Gali}},\ }\bibfield  {title} {\bibinfo {title} {Magneto-optical spectra of the split nickel-vacancy defect in diamond},\ }\href {https://doi.org/10.1103/PhysRevResearch.3.043052} {\bibfield  {journal} {\bibinfo  {journal} {Phys. Rev. Res.}\ }\textbf {\bibinfo {volume} {3}},\ \bibinfo {pages} {043052} (\bibinfo {year} {2021})}\BibitemShut {NoStop}%
\bibitem [{\citenamefont {Goss}\ \emph {et~al.}(2005)\citenamefont {Goss}, \citenamefont {Briddon}, \citenamefont {Rayson}, \citenamefont {Sque},\ and\ \citenamefont {Jones}}]{GossJP:2005PRB}%
  \BibitemOpen
  \bibfield  {author} {\bibinfo {author} {\bibfnamefont {J.~P.}\ \bibnamefont {Goss}}, \bibinfo {author} {\bibfnamefont {P.~R.}\ \bibnamefont {Briddon}}, \bibinfo {author} {\bibfnamefont {M.~J.}\ \bibnamefont {Rayson}}, \bibinfo {author} {\bibfnamefont {S.~J.}\ \bibnamefont {Sque}},\ and\ \bibinfo {author} {\bibfnamefont {R.}~\bibnamefont {Jones}},\ }\bibfield  {title} {\bibinfo {title} {Vacancy-impurity complexes and limitations for implantation doping of diamond},\ }\href {https://doi.org/10.1103/PhysRevB.72.035214} {\bibfield  {journal} {\bibinfo  {journal} {Phys. Rev. B}\ }\textbf {\bibinfo {volume} {72}},\ \bibinfo {pages} {035214} (\bibinfo {year} {2005})}\BibitemShut {NoStop}%
\bibitem [{\citenamefont {Vanpoucke}\ and\ \citenamefont {Haenen}(2017)}]{VanpouckeDEP:DiamRelatMater2017}%
  \BibitemOpen
  \bibfield  {author} {\bibinfo {author} {\bibfnamefont {D.~E.~P.}\ \bibnamefont {Vanpoucke}}\ and\ \bibinfo {author} {\bibfnamefont {K.}~\bibnamefont {Haenen}},\ }\bibfield  {title} {\bibinfo {title} {Revisiting the neutral {C}-vacancy in diamond: {L}ocalization of electrons through {DFT+U}},\ }\href {https://doi.org/10.1016/j.diamond.2017.08.009} {\bibfield  {journal} {\bibinfo  {journal} {Diam. Relat. Mater.}\ }\textbf {\bibinfo {volume} {79}},\ \bibinfo {pages} {60} (\bibinfo {year} {2017})}\BibitemShut {NoStop}%
\bibitem [{\citenamefont {L\"{o}fgren}\ \emph {et~al.}(2018)\citenamefont {L\"{o}fgren}, \citenamefont {Pawar}, \citenamefont {\"{O}berg},\ and\ \citenamefont {Larsson}}]{LofgrenR:NewJPhys2018}%
  \BibitemOpen
  \bibfield  {author} {\bibinfo {author} {\bibfnamefont {R.}~\bibnamefont {L\"{o}fgren}}, \bibinfo {author} {\bibfnamefont {R.}~\bibnamefont {Pawar}}, \bibinfo {author} {\bibfnamefont {S.}~\bibnamefont {\"{O}berg}},\ and\ \bibinfo {author} {\bibfnamefont {J.~A.}\ \bibnamefont {Larsson}},\ }\bibfield  {title} {\bibinfo {title} {Charged dopants in neutral supercells through substitutional donor (acceptor): nitrogen donor charging of the nitrogen-vacancy center in diamond},\ }\href {https://doi.org/10.1088/1367-2630/aaa382} {\bibfield  {journal} {\bibinfo  {journal} {New J. Phys.}\ }\textbf {\bibinfo {volume} {20}},\ \bibinfo {pages} {023002} (\bibinfo {year} {2018})}\BibitemShut {NoStop}%
\bibitem [{\citenamefont {Lindner}\ \emph {et~al.}(2018)\citenamefont {Lindner}, \citenamefont {Bommer}, \citenamefont {Muzha}, \citenamefont {Krueger}, \citenamefont {Gines}, \citenamefont {Mandal}, \citenamefont {Williams}, \citenamefont {Londero}, \citenamefont {Gali},\ and\ \citenamefont {Becher}}]{lindner:2018njp}%
  \BibitemOpen
  \bibfield  {author} {\bibinfo {author} {\bibfnamefont {S.}~\bibnamefont {Lindner}}, \bibinfo {author} {\bibfnamefont {A.}~\bibnamefont {Bommer}}, \bibinfo {author} {\bibfnamefont {A.}~\bibnamefont {Muzha}}, \bibinfo {author} {\bibfnamefont {A.}~\bibnamefont {Krueger}}, \bibinfo {author} {\bibfnamefont {L.}~\bibnamefont {Gines}}, \bibinfo {author} {\bibfnamefont {S.}~\bibnamefont {Mandal}}, \bibinfo {author} {\bibfnamefont {O.}~\bibnamefont {Williams}}, \bibinfo {author} {\bibfnamefont {E.}~\bibnamefont {Londero}}, \bibinfo {author} {\bibfnamefont {A.}~\bibnamefont {Gali}},\ and\ \bibinfo {author} {\bibfnamefont {C.}~\bibnamefont {Becher}},\ }\bibfield  {title} {\bibinfo {title} {{Strongly inhomogeneous distribution of spectral properties of silicon-vacancy color centers in nanodiamonds}},\ }\href {https://doi.org/10.1088/1367-2630/aae93f} {\bibfield  {journal} {\bibinfo  {journal} {New J. Phys.}\ }\textbf {\bibinfo {volume} {20}},\ \bibinfo {pages} {115002} (\bibinfo {year} {2018})}\BibitemShut {NoStop}%
\bibitem [{\citenamefont {Ekimov}\ \emph {et~al.}(2019)\citenamefont {Ekimov}, \citenamefont {Lyapin}, \citenamefont {Razgulov},\ and\ \citenamefont {Kondrin}}]{ekimovE:2019jetp}%
  \BibitemOpen
  \bibfield  {author} {\bibinfo {author} {\bibfnamefont {E.~A.}\ \bibnamefont {Ekimov}}, \bibinfo {author} {\bibfnamefont {S.~G.}\ \bibnamefont {Lyapin}}, \bibinfo {author} {\bibfnamefont {A.~A.}\ \bibnamefont {Razgulov}},\ and\ \bibinfo {author} {\bibfnamefont {M.~V.}\ \bibnamefont {Kondrin}},\ }\bibfield  {title} {\bibinfo {title} {{Ab initio Calculation of Impurity–Vacancy Complexes in Diamond at High Pressure}},\ }\href {https://doi.org/10.1134/s1063776119090097} {\bibfield  {journal} {\bibinfo  {journal} {J. Exp. Theor. Phys.}\ }\textbf {\bibinfo {volume} {129}},\ \bibinfo {pages} {855} (\bibinfo {year} {2019})}\BibitemShut {NoStop}%
\bibitem [{\citenamefont {Vanpoucke}\ \emph {et~al.}(2019)\citenamefont {Vanpoucke}, \citenamefont {Nicley}, \citenamefont {Raymakers}, \citenamefont {Maes},\ and\ \citenamefont {Haenen}}]{VanpouckeDEP:DiamRelatMater2019}%
  \BibitemOpen
  \bibfield  {author} {\bibinfo {author} {\bibfnamefont {D.~E.~P.}\ \bibnamefont {Vanpoucke}}, \bibinfo {author} {\bibfnamefont {S.~S.}\ \bibnamefont {Nicley}}, \bibinfo {author} {\bibfnamefont {J.}~\bibnamefont {Raymakers}}, \bibinfo {author} {\bibfnamefont {W.}~\bibnamefont {Maes}},\ and\ \bibinfo {author} {\bibfnamefont {K.}~\bibnamefont {Haenen}},\ }\bibfield  {title} {\bibinfo {title} {Can europium atoms form luminescent centres in diamond: {A} combined theoretical--experimental study},\ }\href {https://doi.org/10.1016/j.diamond.2019.02.024} {\bibfield  {journal} {\bibinfo  {journal} {Diam. Relat. Mater.}\ }\textbf {\bibinfo {volume} {94}},\ \bibinfo {pages} {233} (\bibinfo {year} {2019})}\BibitemShut {NoStop}%
\bibitem [{\citenamefont {Doherty}\ \emph {et~al.}(2013)\citenamefont {Doherty}, \citenamefont {Manson}, \citenamefont {Delaney}, \citenamefont {Jelezko}, \citenamefont {Wrachtrup},\ and\ \citenamefont {Hollenberg}}]{DOHERTY:2013pr}%
  \BibitemOpen
  \bibfield  {author} {\bibinfo {author} {\bibfnamefont {M.~W.}\ \bibnamefont {Doherty}}, \bibinfo {author} {\bibfnamefont {N.~B.}\ \bibnamefont {Manson}}, \bibinfo {author} {\bibfnamefont {P.}~\bibnamefont {Delaney}}, \bibinfo {author} {\bibfnamefont {F.}~\bibnamefont {Jelezko}}, \bibinfo {author} {\bibfnamefont {J.}~\bibnamefont {Wrachtrup}},\ and\ \bibinfo {author} {\bibfnamefont {L.~C.~L.}\ \bibnamefont {Hollenberg}},\ }\bibfield  {title} {\bibinfo {title} {The nitrogen-vacancy colour centre in diamond},\ }\href {https://doi.org/https://doi.org/10.1016/j.physrep.2013.02.001} {\bibfield  {journal} {\bibinfo  {journal} {Phys. Rep.}\ }\textbf {\bibinfo {volume} {528}},\ \bibinfo {pages} {1} (\bibinfo {year} {2013})}\BibitemShut {NoStop}%
\bibitem [{\citenamefont {Neu}\ \emph {et~al.}(2011)\citenamefont {Neu}, \citenamefont {Fischer}, \citenamefont {Gsell}, \citenamefont {Schreck},\ and\ \citenamefont {Becher}}]{NeuE:2011prb}%
  \BibitemOpen
  \bibfield  {author} {\bibinfo {author} {\bibfnamefont {E.}~\bibnamefont {Neu}}, \bibinfo {author} {\bibfnamefont {M.}~\bibnamefont {Fischer}}, \bibinfo {author} {\bibfnamefont {S.}~\bibnamefont {Gsell}}, \bibinfo {author} {\bibfnamefont {M.}~\bibnamefont {Schreck}},\ and\ \bibinfo {author} {\bibfnamefont {C.}~\bibnamefont {Becher}},\ }\bibfield  {title} {\bibinfo {title} {Fluorescence and polarization spectroscopy of single silicon vacancy centers in heteroepitaxial nanodiamonds on iridium},\ }\href {https://doi.org/10.1103/PhysRevB.84.205211} {\bibfield  {journal} {\bibinfo  {journal} {Phys. Rev. B}\ }\textbf {\bibinfo {volume} {84}},\ \bibinfo {pages} {205211} (\bibinfo {year} {2011})}\BibitemShut {NoStop}%
\bibitem [{\citenamefont {Siyushev}\ \emph {et~al.}(2017)\citenamefont {Siyushev}, \citenamefont {Metsch}, \citenamefont {Ijaz}, \citenamefont {Binder}, \citenamefont {Bhaskar}, \citenamefont {Sukachev}, \citenamefont {Sipahigil}, \citenamefont {Evans}, \citenamefont {Nguyen}, \citenamefont {Lukin}, \citenamefont {Hemmer}, \citenamefont {Palyanov}, \citenamefont {Kupriyanov}, \citenamefont {Borzdov}, \citenamefont {Rogers},\ and\ \citenamefont {Jelezko}}]{Siyushev:2017prb}%
  \BibitemOpen
  \bibfield  {author} {\bibinfo {author} {\bibfnamefont {P.}~\bibnamefont {Siyushev}}, \bibinfo {author} {\bibfnamefont {M.~H.}\ \bibnamefont {Metsch}}, \bibinfo {author} {\bibfnamefont {A.}~\bibnamefont {Ijaz}}, \bibinfo {author} {\bibfnamefont {J.~M.}\ \bibnamefont {Binder}}, \bibinfo {author} {\bibfnamefont {M.~K.}\ \bibnamefont {Bhaskar}}, \bibinfo {author} {\bibfnamefont {D.~D.}\ \bibnamefont {Sukachev}}, \bibinfo {author} {\bibfnamefont {A.}~\bibnamefont {Sipahigil}}, \bibinfo {author} {\bibfnamefont {R.~E.}\ \bibnamefont {Evans}}, \bibinfo {author} {\bibfnamefont {C.~T.}\ \bibnamefont {Nguyen}}, \bibinfo {author} {\bibfnamefont {M.~D.}\ \bibnamefont {Lukin}}, \bibinfo {author} {\bibfnamefont {P.~R.}\ \bibnamefont {Hemmer}}, \bibinfo {author} {\bibfnamefont {Y.~N.}\ \bibnamefont {Palyanov}}, \bibinfo {author} {\bibfnamefont {I.~N.}\ \bibnamefont {Kupriyanov}}, \bibinfo {author} {\bibfnamefont {Y.~M.}\ \bibnamefont {Borzdov}}, \bibinfo {author} {\bibfnamefont {L.~J.}\ \bibnamefont {Rogers}},\ and\
  \bibinfo {author} {\bibfnamefont {F.}~\bibnamefont {Jelezko}},\ }\bibfield  {title} {\bibinfo {title} {Optical and microwave control of germanium-vacancy center spins in diamond},\ }\href {https://doi.org/10.1103/PhysRevB.96.081201} {\bibfield  {journal} {\bibinfo  {journal} {Phys. Rev. B}\ }\textbf {\bibinfo {volume} {96}},\ \bibinfo {pages} {081201} (\bibinfo {year} {2017})}\BibitemShut {NoStop}%
\bibitem [{\citenamefont {Iwasaki}\ \emph {et~al.}(2015)\citenamefont {Iwasaki}, \citenamefont {Ishibashi}, \citenamefont {Miyamoto}, \citenamefont {Doi}, \citenamefont {Kobayashi}, \citenamefont {Miyazaki}, \citenamefont {Tahara}, \citenamefont {Jahnke}, \citenamefont {Rogers}, \citenamefont {Naydenov}, \citenamefont {Jelezko}, \citenamefont {Yamasaki}, \citenamefont {Nagamachi}, \citenamefont {Inubushi}, \citenamefont {Mizuochi},\ and\ \citenamefont {Hatano}}]{iwasaki:2015SR}%
  \BibitemOpen
  \bibfield  {author} {\bibinfo {author} {\bibfnamefont {T.}~\bibnamefont {Iwasaki}}, \bibinfo {author} {\bibfnamefont {F.}~\bibnamefont {Ishibashi}}, \bibinfo {author} {\bibfnamefont {Y.}~\bibnamefont {Miyamoto}}, \bibinfo {author} {\bibfnamefont {Y.}~\bibnamefont {Doi}}, \bibinfo {author} {\bibfnamefont {S.}~\bibnamefont {Kobayashi}}, \bibinfo {author} {\bibfnamefont {T.}~\bibnamefont {Miyazaki}}, \bibinfo {author} {\bibfnamefont {K.}~\bibnamefont {Tahara}}, \bibinfo {author} {\bibfnamefont {K.~D.}\ \bibnamefont {Jahnke}}, \bibinfo {author} {\bibfnamefont {L.~J.}\ \bibnamefont {Rogers}}, \bibinfo {author} {\bibfnamefont {B.}~\bibnamefont {Naydenov}}, \bibinfo {author} {\bibfnamefont {F.}~\bibnamefont {Jelezko}}, \bibinfo {author} {\bibfnamefont {S.}~\bibnamefont {Yamasaki}}, \bibinfo {author} {\bibfnamefont {S.}~\bibnamefont {Nagamachi}}, \bibinfo {author} {\bibfnamefont {T.}~\bibnamefont {Inubushi}}, \bibinfo {author} {\bibfnamefont {N.}~\bibnamefont {Mizuochi}},\ and\ \bibinfo {author} {\bibfnamefont
  {M.}~\bibnamefont {Hatano}},\ }\bibfield  {title} {\bibinfo {title} {{Germanium-Vacancy single color centers in Diamond}},\ }\href {https://doi.org/10.1038/srep12882} {\bibfield  {journal} {\bibinfo  {journal} {Sci. Rep}\ }\textbf {\bibinfo {volume} {5}},\ \bibinfo {pages} {12882} (\bibinfo {year} {2015})}\BibitemShut {NoStop}%
\bibitem [{\citenamefont {Iwasaki}\ \emph {et~al.}(2017)\citenamefont {Iwasaki}, \citenamefont {Miyamoto}, \citenamefont {Taniguchi}, \citenamefont {Siyushev}, \citenamefont {Metsch}, \citenamefont {Jelezko},\ and\ \citenamefont {Hatano}}]{Iwasaki:2017PRL_SnV}%
  \BibitemOpen
  \bibfield  {author} {\bibinfo {author} {\bibfnamefont {T.}~\bibnamefont {Iwasaki}}, \bibinfo {author} {\bibfnamefont {Y.}~\bibnamefont {Miyamoto}}, \bibinfo {author} {\bibfnamefont {T.}~\bibnamefont {Taniguchi}}, \bibinfo {author} {\bibfnamefont {P.}~\bibnamefont {Siyushev}}, \bibinfo {author} {\bibfnamefont {M.~H.}\ \bibnamefont {Metsch}}, \bibinfo {author} {\bibfnamefont {F.}~\bibnamefont {Jelezko}},\ and\ \bibinfo {author} {\bibfnamefont {M.}~\bibnamefont {Hatano}},\ }\bibfield  {title} {\bibinfo {title} {{Tin-Vacancy Quantum Emitters in Diamond}},\ }\href {https://doi.org/10.1103/PhysRevLett.119.253601} {\bibfield  {journal} {\bibinfo  {journal} {Phys. Rev. Lett.}\ }\textbf {\bibinfo {volume} {119}},\ \bibinfo {pages} {253601} (\bibinfo {year} {2017})}\BibitemShut {NoStop}%
\bibitem [{\citenamefont {Wang}\ \emph {et~al.}(2024)\citenamefont {Wang}, \citenamefont {Kazak}, \citenamefont {Senkalla}, \citenamefont {Siyushev}, \citenamefont {Abe}, \citenamefont {Taniguchi}, \citenamefont {Onoda}, \citenamefont {Kato}, \citenamefont {Makino}, \citenamefont {Hatano}, \citenamefont {Jelezko},\ and\ \citenamefont {Iwasaki}}]{Wang:2024PhysRevLett_PbV}%
  \BibitemOpen
  \bibfield  {author} {\bibinfo {author} {\bibfnamefont {P.}~\bibnamefont {Wang}}, \bibinfo {author} {\bibfnamefont {L.}~\bibnamefont {Kazak}}, \bibinfo {author} {\bibfnamefont {K.}~\bibnamefont {Senkalla}}, \bibinfo {author} {\bibfnamefont {P.}~\bibnamefont {Siyushev}}, \bibinfo {author} {\bibfnamefont {R.}~\bibnamefont {Abe}}, \bibinfo {author} {\bibfnamefont {T.}~\bibnamefont {Taniguchi}}, \bibinfo {author} {\bibfnamefont {S.}~\bibnamefont {Onoda}}, \bibinfo {author} {\bibfnamefont {H.}~\bibnamefont {Kato}}, \bibinfo {author} {\bibfnamefont {T.}~\bibnamefont {Makino}}, \bibinfo {author} {\bibfnamefont {M.}~\bibnamefont {Hatano}}, \bibinfo {author} {\bibfnamefont {F.}~\bibnamefont {Jelezko}},\ and\ \bibinfo {author} {\bibfnamefont {T.}~\bibnamefont {Iwasaki}},\ }\bibfield  {title} {\bibinfo {title} {{Transform-Limited Photon Emission from a Lead-Vacancy Center in Diamond above 10 K}},\ }\href {https://doi.org/10.1103/PhysRevLett.132.073601} {\bibfield  {journal} {\bibinfo  {journal} {Phys. Rev. Lett.}\
  }\textbf {\bibinfo {volume} {132}},\ \bibinfo {pages} {073601} (\bibinfo {year} {2024})}\BibitemShut {NoStop}%
\bibitem [{\citenamefont {Abe}\ \emph {et~al.}(2026)\citenamefont {Abe}, \citenamefont {Chen}, \citenamefont {Wang}, \citenamefont {Taniguchi}, \citenamefont {Miyakawa}, \citenamefont {Onoda}, \citenamefont {Hatano},\ and\ \citenamefont {Iwasaki}}]{Abe:2026AdvFunctMat_PbV}%
  \BibitemOpen
  \bibfield  {author} {\bibinfo {author} {\bibfnamefont {R.}~\bibnamefont {Abe}}, \bibinfo {author} {\bibfnamefont {Y.}~\bibnamefont {Chen}}, \bibinfo {author} {\bibfnamefont {P.}~\bibnamefont {Wang}}, \bibinfo {author} {\bibfnamefont {T.}~\bibnamefont {Taniguchi}}, \bibinfo {author} {\bibfnamefont {M.}~\bibnamefont {Miyakawa}}, \bibinfo {author} {\bibfnamefont {S.}~\bibnamefont {Onoda}}, \bibinfo {author} {\bibfnamefont {M.}~\bibnamefont {Hatano}},\ and\ \bibinfo {author} {\bibfnamefont {T.}~\bibnamefont {Iwasaki}},\ }\bibfield  {title} {\bibinfo {title} {{Narrow Inhomogeneous Distribution and Charge State Stabilization of Lead-Vacancy Centers in Diamond}},\ }\href {https://doi.org/10.1002/adfm.202512412} {\bibfield  {journal} {\bibinfo  {journal} {Adv. Funct. Mater.}\ }\textbf {\bibinfo {volume} {36}},\ \bibinfo {pages} {e12412} (\bibinfo {year} {2026})}\BibitemShut {NoStop}%
\bibitem [{\citenamefont {Thiering}\ and\ \citenamefont {Gali}(2018)}]{thiering:2018prx}%
  \BibitemOpen
  \bibfield  {author} {\bibinfo {author} {\bibfnamefont {G.}~\bibnamefont {Thiering}}\ and\ \bibinfo {author} {\bibfnamefont {A.}~\bibnamefont {Gali}},\ }\bibfield  {title} {\bibinfo {title} {{Ab initio Magneto-Optical Spectrum of Group-IV vacancy Color centers in Diamond}},\ }\href {https://doi.org/10.1103/physrevx.8.021063} {\bibfield  {journal} {\bibinfo  {journal} {Phys. Rev. X}\ }\textbf {\bibinfo {volume} {8}},\ \bibinfo {pages} {021063} (\bibinfo {year} {2018})}\BibitemShut {NoStop}%
\bibitem [{\citenamefont {Wahl}\ \emph {et~al.}(2020)\citenamefont {Wahl}, \citenamefont {Correia}, \citenamefont {Villarreal}, \citenamefont {Bourgeois}, \citenamefont {Gulka}, \citenamefont {Nesl\'adek}, \citenamefont {Vantomme},\ and\ \citenamefont {Pereira}}]{WahlPereira:PhysRevLett2020_SnV}%
  \BibitemOpen
  \bibfield  {author} {\bibinfo {author} {\bibfnamefont {U.}~\bibnamefont {Wahl}}, \bibinfo {author} {\bibfnamefont {J.~G.}\ \bibnamefont {Correia}}, \bibinfo {author} {\bibfnamefont {R.}~\bibnamefont {Villarreal}}, \bibinfo {author} {\bibfnamefont {E.}~\bibnamefont {Bourgeois}}, \bibinfo {author} {\bibfnamefont {M.}~\bibnamefont {Gulka}}, \bibinfo {author} {\bibfnamefont {M.}~\bibnamefont {Nesl\'adek}}, \bibinfo {author} {\bibfnamefont {A.}~\bibnamefont {Vantomme}},\ and\ \bibinfo {author} {\bibfnamefont {L.~M.~C.}\ \bibnamefont {Pereira}},\ }\bibfield  {title} {\bibinfo {title} {{Direct Structural Identification and Quantification of the Split-Vacancy Configuration for Implanted Sn in Diamond}},\ }\href {https://doi.org/10.1103/PhysRevLett.125.045301} {\bibfield  {journal} {\bibinfo  {journal} {Phys. Rev. Lett.}\ }\textbf {\bibinfo {volume} {125}},\ \bibinfo {pages} {045301} (\bibinfo {year} {2020})}\BibitemShut {NoStop}%
\bibitem [{\citenamefont {Chen}\ \emph {et~al.}(2020)\citenamefont {Chen}, \citenamefont {Zheludev},\ and\ \citenamefont {Gao}}]{chen:2020aqt}%
  \BibitemOpen
  \bibfield  {author} {\bibinfo {author} {\bibfnamefont {D.}~\bibnamefont {Chen}}, \bibinfo {author} {\bibfnamefont {N.}~\bibnamefont {Zheludev}},\ and\ \bibinfo {author} {\bibfnamefont {W.}~\bibnamefont {Gao}},\ }\bibfield  {title} {\bibinfo {title} {{Building blocks for quantum network based on Group‐IV Split‐Vacancy centers in Diamond}},\ }\href {https://doi.org/10.1002/qute.201900069} {\bibfield  {journal} {\bibinfo  {journal} {Adv. Quantum Technol}\ }\textbf {\bibinfo {volume} {3}},\ \bibinfo {pages} {1900069} (\bibinfo {year} {2020})}\BibitemShut {NoStop}%
\bibitem [{\citenamefont {Karapatzakis}\ \emph {et~al.}(2024)\citenamefont {Karapatzakis}, \citenamefont {Resch}, \citenamefont {Schrodin}, \citenamefont {Fuchs}, \citenamefont {Kieschnick}, \citenamefont {Heupel}, \citenamefont {Kussi}, \citenamefont {S\"urgers}, \citenamefont {Popov}, \citenamefont {Meijer}, \citenamefont {Becher}, \citenamefont {Wernsdorfer},\ and\ \citenamefont {Hunger}}]{Karapatzakis:2024PhysRevX_GSsplit}%
  \BibitemOpen
  \bibfield  {author} {\bibinfo {author} {\bibfnamefont {I.}~\bibnamefont {Karapatzakis}}, \bibinfo {author} {\bibfnamefont {J.}~\bibnamefont {Resch}}, \bibinfo {author} {\bibfnamefont {M.}~\bibnamefont {Schrodin}}, \bibinfo {author} {\bibfnamefont {P.}~\bibnamefont {Fuchs}}, \bibinfo {author} {\bibfnamefont {M.}~\bibnamefont {Kieschnick}}, \bibinfo {author} {\bibfnamefont {J.}~\bibnamefont {Heupel}}, \bibinfo {author} {\bibfnamefont {L.}~\bibnamefont {Kussi}}, \bibinfo {author} {\bibfnamefont {C.}~\bibnamefont {S\"urgers}}, \bibinfo {author} {\bibfnamefont {C.}~\bibnamefont {Popov}}, \bibinfo {author} {\bibfnamefont {J.}~\bibnamefont {Meijer}}, \bibinfo {author} {\bibfnamefont {C.}~\bibnamefont {Becher}}, \bibinfo {author} {\bibfnamefont {W.}~\bibnamefont {Wernsdorfer}},\ and\ \bibinfo {author} {\bibfnamefont {D.}~\bibnamefont {Hunger}},\ }\bibfield  {title} {\bibinfo {title} {{Microwave Control of the Tin-Vacancy Spin Qubit in Diamond with a Superconducting Waveguide}},\ }\href
  {https://doi.org/10.1103/PhysRevX.14.031036} {\bibfield  {journal} {\bibinfo  {journal} {Phys. Rev. X}\ }\textbf {\bibinfo {volume} {14}},\ \bibinfo {pages} {031036} (\bibinfo {year} {2024})}\BibitemShut {NoStop}%
\bibitem [{\citenamefont {Mary~Joy}\ \emph {et~al.}(2026)\citenamefont {Mary~Joy}, \citenamefont {Cherta~Garrido}, \citenamefont {Harb}, \citenamefont {Jeuris}, \citenamefont {Rouzbahani}, \citenamefont {D'Haen}, \citenamefont {Clemmen}, \citenamefont {Van~Thourhout}, \citenamefont {Vanpoucke}, \citenamefont {Pobedinskas},\ and\ \citenamefont {Haenen}}]{MaryJoyVanpoucke:ACSMaterSci2026}%
  \BibitemOpen
  \bibfield  {author} {\bibinfo {author} {\bibfnamefont {R.}~\bibnamefont {Mary~Joy}}, \bibinfo {author} {\bibfnamefont {M.}~\bibnamefont {Cherta~Garrido}}, \bibinfo {author} {\bibfnamefont {O.~J.~Y.}\ \bibnamefont {Harb}}, \bibinfo {author} {\bibfnamefont {H.}~\bibnamefont {Jeuris}}, \bibinfo {author} {\bibfnamefont {R.}~\bibnamefont {Rouzbahani}}, \bibinfo {author} {\bibfnamefont {J.}~\bibnamefont {D'Haen}}, \bibinfo {author} {\bibfnamefont {S.}~\bibnamefont {Clemmen}}, \bibinfo {author} {\bibfnamefont {D.}~\bibnamefont {Van~Thourhout}}, \bibinfo {author} {\bibfnamefont {D.~E.~P.}\ \bibnamefont {Vanpoucke}}, \bibinfo {author} {\bibfnamefont {P.}~\bibnamefont {Pobedinskas}},\ and\ \bibinfo {author} {\bibfnamefont {K.}~\bibnamefont {Haenen}},\ }\bibfield  {title} {\bibinfo {title} {{Fabrication and Photoluminescence Studies of Tin-Vacancy Centers in Chemical Vapor Deposition Diamond}},\ }\href {https://doi.org/10.1021/acsmaterialslett.5c01218} {\bibfield  {journal} {\bibinfo  {journal} {ACS Materials Lett.}\
  }\textbf {\bibinfo {volume} {8}},\ \bibinfo {pages} {137} (\bibinfo {year} {2026})}\BibitemShut {NoStop}%
\bibitem [{\citenamefont {G\"{o}rlitz}\ \emph {et~al.}(2020)\citenamefont {G\"{o}rlitz}, \citenamefont {Herrmann}, \citenamefont {Thiering}, \citenamefont {Fuchs}, \citenamefont {Gandil}, \citenamefont {Iwasaki}, \citenamefont {Taniguchi}, \citenamefont {Kieschnick}, \citenamefont {Meijer}, \citenamefont {Hatano}, \citenamefont {Gali},\ and\ \citenamefont {Becher}}]{Gorlitz:2020NJPhys}%
  \BibitemOpen
  \bibfield  {author} {\bibinfo {author} {\bibfnamefont {J.}~\bibnamefont {G\"{o}rlitz}}, \bibinfo {author} {\bibfnamefont {D.}~\bibnamefont {Herrmann}}, \bibinfo {author} {\bibfnamefont {G.}~\bibnamefont {Thiering}}, \bibinfo {author} {\bibfnamefont {P.}~\bibnamefont {Fuchs}}, \bibinfo {author} {\bibfnamefont {M.}~\bibnamefont {Gandil}}, \bibinfo {author} {\bibfnamefont {T.}~\bibnamefont {Iwasaki}}, \bibinfo {author} {\bibfnamefont {T.}~\bibnamefont {Taniguchi}}, \bibinfo {author} {\bibfnamefont {M.}~\bibnamefont {Kieschnick}}, \bibinfo {author} {\bibfnamefont {J.}~\bibnamefont {Meijer}}, \bibinfo {author} {\bibfnamefont {M.}~\bibnamefont {Hatano}}, \bibinfo {author} {\bibfnamefont {A.}~\bibnamefont {Gali}},\ and\ \bibinfo {author} {\bibfnamefont {C.}~\bibnamefont {Becher}},\ }\bibfield  {title} {\bibinfo {title} {Spectroscopic investigations of negatively charged tin-vacancy centres in diamond},\ }\href {https://doi.org/10.1088/1367-2630/ab6631} {\bibfield  {journal} {\bibinfo  {journal} {New J. Phys.}\
  }\textbf {\bibinfo {volume} {22}},\ \bibinfo {pages} {013048} (\bibinfo {year} {2020})}\BibitemShut {NoStop}%
\bibitem [{\citenamefont {Ekimov}\ \emph {et~al.}(2018)\citenamefont {Ekimov}, \citenamefont {Lyapin},\ and\ \citenamefont {Kondrin}}]{ekimov:2018DRM}%
  \BibitemOpen
  \bibfield  {author} {\bibinfo {author} {\bibfnamefont {E.~A.}\ \bibnamefont {Ekimov}}, \bibinfo {author} {\bibfnamefont {S.~G.}\ \bibnamefont {Lyapin}},\ and\ \bibinfo {author} {\bibfnamefont {M.~V.}\ \bibnamefont {Kondrin}},\ }\bibfield  {title} {\bibinfo {title} {Tin-vacancy color centers in micro- and polycrystalline diamonds synthesized at high pressures},\ }\href {https://doi.org/10.1016/j.diamond.2018.06.014} {\bibfield  {journal} {\bibinfo  {journal} {Diam. Relat. Mater.}\ }\textbf {\bibinfo {volume} {87}},\ \bibinfo {pages} {223} (\bibinfo {year} {2018})}\BibitemShut {NoStop}%
\bibitem [{\citenamefont {Brevoord}(2025)}]{Brevoord:2025Thesis_SnVexp}%
  \BibitemOpen
  \bibfield  {author} {\bibinfo {author} {\bibfnamefont {J.~M.}\ \bibnamefont {Brevoord}},\ }{\selectlanguage {English}\emph {\bibinfo {title} {Control of the optical interface of color centers in diamond}}},\ \href {https://doi.org/10.4233/uuid:de64a49e-9cfb-4865-a5ce-f506cafd4460} {\bibinfo {type} {Dissertation (tu delft)}},\ \bibinfo  {school} {Delft University of Technology} (\bibinfo {year} {2025})\BibitemShut {NoStop}%
\bibitem [{\citenamefont {Brevoord}\ \emph {et~al.}(2025)\citenamefont {Brevoord}, \citenamefont {Wienhoven}, \citenamefont {Codreanu}, \citenamefont {Ishiguro}, \citenamefont {van Leeuwen}, \citenamefont {Iuliano}, \citenamefont {De~Santis}, \citenamefont {Waas}, \citenamefont {Beukers}, \citenamefont {Turan}, \citenamefont {Errando-Herranz}, \citenamefont {Kawaguchi},\ and\ \citenamefont {Hanson}}]{Brevoord:2025ApplPhysLett}%
  \BibitemOpen
  \bibfield  {author} {\bibinfo {author} {\bibfnamefont {J.~M.}\ \bibnamefont {Brevoord}}, \bibinfo {author} {\bibfnamefont {L.~G.~C.}\ \bibnamefont {Wienhoven}}, \bibinfo {author} {\bibfnamefont {N.}~\bibnamefont {Codreanu}}, \bibinfo {author} {\bibfnamefont {T.}~\bibnamefont {Ishiguro}}, \bibinfo {author} {\bibfnamefont {E.}~\bibnamefont {van Leeuwen}}, \bibinfo {author} {\bibfnamefont {M.}~\bibnamefont {Iuliano}}, \bibinfo {author} {\bibfnamefont {L.}~\bibnamefont {De~Santis}}, \bibinfo {author} {\bibfnamefont {C.}~\bibnamefont {Waas}}, \bibinfo {author} {\bibfnamefont {H.~K.~C.}\ \bibnamefont {Beukers}}, \bibinfo {author} {\bibfnamefont {T.}~\bibnamefont {Turan}}, \bibinfo {author} {\bibfnamefont {C.}~\bibnamefont {Errando-Herranz}}, \bibinfo {author} {\bibfnamefont {K.}~\bibnamefont {Kawaguchi}},\ and\ \bibinfo {author} {\bibfnamefont {R.}~\bibnamefont {Hanson}},\ }\bibfield  {title} {\bibinfo {title} {Large-range tuning and stabilization of the optical transition of diamond tin-vacancy centers by in
  situ strain control},\ }\href {https://doi.org/10.1063/5.0251211} {\bibfield  {journal} {\bibinfo  {journal} {Appl. Phys. Lett.}\ }\textbf {\bibinfo {volume} {126}},\ \bibinfo {pages} {174001} (\bibinfo {year} {2025})}\BibitemShut {NoStop}%
\bibitem [{\citenamefont {Li}\ \emph {et~al.}(2024)\citenamefont {Li}, \citenamefont {{De Santis}}, \citenamefont {Harris}, \citenamefont {Chen}, \citenamefont {Gao}, \citenamefont {Christen}, \citenamefont {Choi}, \citenamefont {Trusheim}, \citenamefont {Song}, \citenamefont {Errando-Herranz}, \citenamefont {Du}, \citenamefont {Hu}, \citenamefont {Clark}, \citenamefont {Ibrahim}, \citenamefont {Gilbert}, \citenamefont {Han},\ and\ \citenamefont {Englund}}]{Li:2024Nature_SnVinCMOS}%
  \BibitemOpen
  \bibfield  {author} {\bibinfo {author} {\bibfnamefont {L.}~\bibnamefont {Li}}, \bibinfo {author} {\bibfnamefont {L.}~\bibnamefont {{De Santis}}}, \bibinfo {author} {\bibfnamefont {I.~B.~W.}\ \bibnamefont {Harris}}, \bibinfo {author} {\bibfnamefont {K.~C.}\ \bibnamefont {Chen}}, \bibinfo {author} {\bibfnamefont {Y.}~\bibnamefont {Gao}}, \bibinfo {author} {\bibfnamefont {I.}~\bibnamefont {Christen}}, \bibinfo {author} {\bibfnamefont {H.}~\bibnamefont {Choi}}, \bibinfo {author} {\bibfnamefont {M.}~\bibnamefont {Trusheim}}, \bibinfo {author} {\bibfnamefont {Y.}~\bibnamefont {Song}}, \bibinfo {author} {\bibfnamefont {C.}~\bibnamefont {Errando-Herranz}}, \bibinfo {author} {\bibfnamefont {J.}~\bibnamefont {Du}}, \bibinfo {author} {\bibfnamefont {Y.}~\bibnamefont {Hu}}, \bibinfo {author} {\bibfnamefont {G.}~\bibnamefont {Clark}}, \bibinfo {author} {\bibfnamefont {M.~I.}\ \bibnamefont {Ibrahim}}, \bibinfo {author} {\bibfnamefont {G.}~\bibnamefont {Gilbert}}, \bibinfo {author} {\bibfnamefont {R.}~\bibnamefont
  {Han}},\ and\ \bibinfo {author} {\bibfnamefont {D.}~\bibnamefont {Englund}},\ }\bibfield  {title} {\bibinfo {title} {Heterogeneous integration of spin--photon interfaces with a {CMOS} platform},\ }\href {https://doi.org/10.1038/s41586-024-07371-7} {\bibfield  {journal} {\bibinfo  {journal} {Nature}\ }\textbf {\bibinfo {volume} {630}},\ \bibinfo {pages} {70} (\bibinfo {year} {2024})}\BibitemShut {NoStop}%
\bibitem [{\citenamefont {Sedov}\ \emph {et~al.}(2023)\citenamefont {Sedov}, \citenamefont {Martyanov}, \citenamefont {Neliubov}, \citenamefont {Tiazhelov}, \citenamefont {Savin}, \citenamefont {Eremchev}, \citenamefont {Eremchev}, \citenamefont {Pavlenko}, \citenamefont {Mandal}, \citenamefont {Ralchenko},\ and\ \citenamefont {Naumov}}]{Sedov:2023PhilTransSoc_SnV}%
  \BibitemOpen
  \bibfield  {author} {\bibinfo {author} {\bibfnamefont {V.}~\bibnamefont {Sedov}}, \bibinfo {author} {\bibfnamefont {A.}~\bibnamefont {Martyanov}}, \bibinfo {author} {\bibfnamefont {A.}~\bibnamefont {Neliubov}}, \bibinfo {author} {\bibfnamefont {I.}~\bibnamefont {Tiazhelov}}, \bibinfo {author} {\bibfnamefont {S.}~\bibnamefont {Savin}}, \bibinfo {author} {\bibfnamefont {I.}~\bibnamefont {Eremchev}}, \bibinfo {author} {\bibfnamefont {M.}~\bibnamefont {Eremchev}}, \bibinfo {author} {\bibfnamefont {M.}~\bibnamefont {Pavlenko}}, \bibinfo {author} {\bibfnamefont {S.}~\bibnamefont {Mandal}}, \bibinfo {author} {\bibfnamefont {V.}~\bibnamefont {Ralchenko}},\ and\ \bibinfo {author} {\bibfnamefont {A.}~\bibnamefont {Naumov}},\ }\bibfield  {title} {\bibinfo {title} {{Narrowband photoluminescence of Tin-Vacancy colour centres in Sn-doped chemical vapour deposition diamond microcrystals}},\ }\href {https://doi.org/10.1098/rsta.2023.0167} {\bibfield  {journal} {\bibinfo  {journal} {Philos. Trans. A Math. Phys. Eng. Sci.}\
  }\textbf {\bibinfo {volume} {382}},\ \bibinfo {pages} {20230167} (\bibinfo {year} {2023})}\BibitemShut {NoStop}%
\bibitem [{\citenamefont {Vindolet}\ \emph {et~al.}(2022)\citenamefont {Vindolet}, \citenamefont {Adam}, \citenamefont {Toraille}, \citenamefont {Chipaux}, \citenamefont {Hilberer}, \citenamefont {Dupuy}, \citenamefont {Razinkovas}, \citenamefont {Alkauskas}, \citenamefont {Thiering}, \citenamefont {Gali}, \citenamefont {De~Feudis}, \citenamefont {Ngandeu~Ngambou}, \citenamefont {Achard}, \citenamefont {Tallaire}, \citenamefont {Schmidt}, \citenamefont {Becher},\ and\ \citenamefont {Roch}}]{VindoletB:PRB2022}%
  \BibitemOpen
  \bibfield  {author} {\bibinfo {author} {\bibfnamefont {B.}~\bibnamefont {Vindolet}}, \bibinfo {author} {\bibfnamefont {M.-P.}\ \bibnamefont {Adam}}, \bibinfo {author} {\bibfnamefont {L.}~\bibnamefont {Toraille}}, \bibinfo {author} {\bibfnamefont {M.}~\bibnamefont {Chipaux}}, \bibinfo {author} {\bibfnamefont {A.}~\bibnamefont {Hilberer}}, \bibinfo {author} {\bibfnamefont {G.}~\bibnamefont {Dupuy}}, \bibinfo {author} {\bibfnamefont {L.}~\bibnamefont {Razinkovas}}, \bibinfo {author} {\bibfnamefont {A.}~\bibnamefont {Alkauskas}}, \bibinfo {author} {\bibfnamefont {G.~m.~H.}\ \bibnamefont {Thiering}}, \bibinfo {author} {\bibfnamefont {A.}~\bibnamefont {Gali}}, \bibinfo {author} {\bibfnamefont {M.}~\bibnamefont {De~Feudis}}, \bibinfo {author} {\bibfnamefont {M.~W.}\ \bibnamefont {Ngandeu~Ngambou}}, \bibinfo {author} {\bibfnamefont {J.}~\bibnamefont {Achard}}, \bibinfo {author} {\bibfnamefont {A.}~\bibnamefont {Tallaire}}, \bibinfo {author} {\bibfnamefont {M.}~\bibnamefont {Schmidt}}, \bibinfo {author}
  {\bibfnamefont {C.}~\bibnamefont {Becher}},\ and\ \bibinfo {author} {\bibfnamefont {J.-F. m.~c.}\ \bibnamefont {Roch}},\ }\bibfield  {title} {\bibinfo {title} {{Optical properties of SiV and GeV color centers in nanodiamonds under hydrostatic pressures up to 180 GPa}},\ }\href {https://doi.org/10.1103/PhysRevB.106.214109} {\bibfield  {journal} {\bibinfo  {journal} {Phys. Rev. B}\ }\textbf {\bibinfo {volume} {106}},\ \bibinfo {pages} {214109} (\bibinfo {year} {2022})}\BibitemShut {NoStop}%
\bibitem [{\citenamefont {Corte}\ \emph {et~al.}(2022)\citenamefont {Corte}, \citenamefont {Sachero}, \citenamefont {Ditalia~Tchernij}, \citenamefont {Lühmann}, \citenamefont {Pezzagna}, \citenamefont {Traina}, \citenamefont {Degiovanni}, \citenamefont {Moreva}, \citenamefont {Olivero}, \citenamefont {Meijer}, \citenamefont {Genovese},\ and\ \citenamefont {Forneris}}]{Corte:2021AdvPhotonRes_SnV0}%
  \BibitemOpen
  \bibfield  {author} {\bibinfo {author} {\bibfnamefont {E.}~\bibnamefont {Corte}}, \bibinfo {author} {\bibfnamefont {S.}~\bibnamefont {Sachero}}, \bibinfo {author} {\bibfnamefont {S.}~\bibnamefont {Ditalia~Tchernij}}, \bibinfo {author} {\bibfnamefont {T.}~\bibnamefont {Lühmann}}, \bibinfo {author} {\bibfnamefont {S.}~\bibnamefont {Pezzagna}}, \bibinfo {author} {\bibfnamefont {P.}~\bibnamefont {Traina}}, \bibinfo {author} {\bibfnamefont {I.~P.}\ \bibnamefont {Degiovanni}}, \bibinfo {author} {\bibfnamefont {E.}~\bibnamefont {Moreva}}, \bibinfo {author} {\bibfnamefont {P.}~\bibnamefont {Olivero}}, \bibinfo {author} {\bibfnamefont {J.}~\bibnamefont {Meijer}}, \bibinfo {author} {\bibfnamefont {M.}~\bibnamefont {Genovese}},\ and\ \bibinfo {author} {\bibfnamefont {J.}~\bibnamefont {Forneris}},\ }\bibfield  {title} {\bibinfo {title} {{Spectral Emission Dependence of Tin-Vacancy Centers in Diamond from Thermal Processing and Chemical Functionalization}},\ }\href {https://doi.org/10.1002/adpr.202100148} {\bibfield
  {journal} {\bibinfo  {journal} {Adv. Photon. Res.}\ }\textbf {\bibinfo {volume} {3}},\ \bibinfo {pages} {2100148} (\bibinfo {year} {2022})}\BibitemShut {NoStop}%
\bibitem [{\citenamefont {Tchernij}\ \emph {et~al.}(2017)\citenamefont {Tchernij}, \citenamefont {Herzig}, \citenamefont {Forneris}, \citenamefont {Küpper}, \citenamefont {Pezzagna}, \citenamefont {Traina}, \citenamefont {Moreva}, \citenamefont {Degiovanni}, \citenamefont {Brida}, \citenamefont {Skukan}, \citenamefont {Genovese}, \citenamefont {Jakšić}, \citenamefont {Meijer},\ and\ \citenamefont {Olivero}}]{tchernij:2017acs}%
  \BibitemOpen
  \bibfield  {author} {\bibinfo {author} {\bibfnamefont {S.~D.}\ \bibnamefont {Tchernij}}, \bibinfo {author} {\bibfnamefont {T.}~\bibnamefont {Herzig}}, \bibinfo {author} {\bibfnamefont {J.}~\bibnamefont {Forneris}}, \bibinfo {author} {\bibfnamefont {J.}~\bibnamefont {Küpper}}, \bibinfo {author} {\bibfnamefont {S.}~\bibnamefont {Pezzagna}}, \bibinfo {author} {\bibfnamefont {P.}~\bibnamefont {Traina}}, \bibinfo {author} {\bibfnamefont {E.}~\bibnamefont {Moreva}}, \bibinfo {author} {\bibfnamefont {I.~P.}\ \bibnamefont {Degiovanni}}, \bibinfo {author} {\bibfnamefont {G.}~\bibnamefont {Brida}}, \bibinfo {author} {\bibfnamefont {N.}~\bibnamefont {Skukan}}, \bibinfo {author} {\bibfnamefont {M.}~\bibnamefont {Genovese}}, \bibinfo {author} {\bibfnamefont {M.}~\bibnamefont {Jakšić}}, \bibinfo {author} {\bibfnamefont {J.}~\bibnamefont {Meijer}},\ and\ \bibinfo {author} {\bibfnamefont {P.}~\bibnamefont {Olivero}},\ }\bibfield  {title} {\bibinfo {title} {{Single-Photon-Emitting Optical Centers in Diamond Fabricated
  upon Sn Implantation}},\ }\href {https://doi.org/10.1021/acsphotonics.7b00904} {\bibfield  {journal} {\bibinfo  {journal} {ACS Photonics}\ }\textbf {\bibinfo {volume} {4}},\ \bibinfo {pages} {2580} (\bibinfo {year} {2017})}\BibitemShut {NoStop}%
\bibitem [{\citenamefont {Kresse}\ and\ \citenamefont {Joubert}(1999)}]{KresseG:PhysRevB1999}%
  \BibitemOpen
  \bibfield  {author} {\bibinfo {author} {\bibfnamefont {G.}~\bibnamefont {Kresse}}\ and\ \bibinfo {author} {\bibfnamefont {D.}~\bibnamefont {Joubert}},\ }\bibfield  {title} {\bibinfo {title} {From ultrasoft pseudopotentials to the projector augmented-wave method},\ }\href {https://doi.org/10.1103/PhysRevB.59.1758} {\bibfield  {journal} {\bibinfo  {journal} {Phys. Rev. B}\ }\textbf {\bibinfo {volume} {59}},\ \bibinfo {pages} {1758} (\bibinfo {year} {1999})}\BibitemShut {NoStop}%
\bibitem [{\citenamefont {Kresse}\ and\ \citenamefont {Furthm\"uller}(1996)}]{KresseG:PhysRevB1996}%
  \BibitemOpen
  \bibfield  {author} {\bibinfo {author} {\bibfnamefont {G.}~\bibnamefont {Kresse}}\ and\ \bibinfo {author} {\bibfnamefont {J.}~\bibnamefont {Furthm\"uller}},\ }\bibfield  {title} {\bibinfo {title} {Efficient iterative schemes for ab initio total-energy calculations using a plane-wave basis set},\ }\href {https://doi.org/10.1103/PhysRevB.54.11169} {\bibfield  {journal} {\bibinfo  {journal} {Phys. Rev. B}\ }\textbf {\bibinfo {volume} {54}},\ \bibinfo {pages} {11169} (\bibinfo {year} {1996})}\BibitemShut {NoStop}%
\bibitem [{\citenamefont {Perdew}\ \emph {et~al.}(1996)\citenamefont {Perdew}, \citenamefont {Burke},\ and\ \citenamefont {Ernzerhof}}]{PBE_PerdewJP:PhysRevLett1996}%
  \BibitemOpen
  \bibfield  {author} {\bibinfo {author} {\bibfnamefont {J.~P.}\ \bibnamefont {Perdew}}, \bibinfo {author} {\bibfnamefont {K.}~\bibnamefont {Burke}},\ and\ \bibinfo {author} {\bibfnamefont {M.}~\bibnamefont {Ernzerhof}},\ }\bibfield  {title} {\bibinfo {title} {{Generalized Gradient Approximation Made Simple}},\ }\href {https://doi.org/10.1103/PhysRevLett.77.3865} {\bibfield  {journal} {\bibinfo  {journal} {Phys. Rev. Lett.}\ }\textbf {\bibinfo {volume} {77}},\ \bibinfo {pages} {3865} (\bibinfo {year} {1996})}\BibitemShut {NoStop}%
\bibitem [{\citenamefont {Krukau}\ \emph {et~al.}(2006)\citenamefont {Krukau}, \citenamefont {Vydrov}, \citenamefont {Izmaylov},\ and\ \citenamefont {Scuseria}}]{HSE06paper_KrukauAV:JChemPhys2006}%
  \BibitemOpen
  \bibfield  {author} {\bibinfo {author} {\bibfnamefont {A.~V.}\ \bibnamefont {Krukau}}, \bibinfo {author} {\bibfnamefont {O.~A.}\ \bibnamefont {Vydrov}}, \bibinfo {author} {\bibfnamefont {A.~F.}\ \bibnamefont {Izmaylov}},\ and\ \bibinfo {author} {\bibfnamefont {G.~E.}\ \bibnamefont {Scuseria}},\ }\bibfield  {title} {\bibinfo {title} {Influence of the exchange screening parameter on the performance of screened hybrid functionals},\ }\href {https://doi.org/10.1063/1.2404663} {\bibfield  {journal} {\bibinfo  {journal} {J Chem. Phys.}\ }\textbf {\bibinfo {volume} {125}},\ \bibinfo {pages} {224106} (\bibinfo {year} {2006})}\BibitemShut {NoStop}%
\bibitem [{\citenamefont {Vinet}\ \emph {et~al.}(1987)\citenamefont {Vinet}, \citenamefont {Smith}, \citenamefont {Ferrante},\ and\ \citenamefont {Rose}}]{VinetP:PhysRevB1987_RoseVinetEOS}%
  \BibitemOpen
  \bibfield  {author} {\bibinfo {author} {\bibfnamefont {P.}~\bibnamefont {Vinet}}, \bibinfo {author} {\bibfnamefont {J.~R.}\ \bibnamefont {Smith}}, \bibinfo {author} {\bibfnamefont {J.}~\bibnamefont {Ferrante}},\ and\ \bibinfo {author} {\bibfnamefont {J.~H.}\ \bibnamefont {Rose}},\ }\bibfield  {title} {\bibinfo {title} {Temperature effects on the universal equation of state of solids},\ }\href {https://doi.org/10.1103/PhysRevB.35.1945} {\bibfield  {journal} {\bibinfo  {journal} {Phys. Rev. B}\ }\textbf {\bibinfo {volume} {35}},\ \bibinfo {pages} {1945} (\bibinfo {year} {1987})}\BibitemShut {NoStop}%
\bibitem [{\citenamefont {Vanpoucke}\ \emph {et~al.}(2013{\natexlab{a}})\citenamefont {Vanpoucke}, \citenamefont {Bultinck},\ and\ \citenamefont {Van~Driessche}}]{VanpouckeDannyEP:2013aJComputChem}%
  \BibitemOpen
  \bibfield  {author} {\bibinfo {author} {\bibfnamefont {D.~E.~P.}\ \bibnamefont {Vanpoucke}}, \bibinfo {author} {\bibfnamefont {P.}~\bibnamefont {Bultinck}},\ and\ \bibinfo {author} {\bibfnamefont {I.}~\bibnamefont {Van~Driessche}},\ }\bibfield  {title} {\bibinfo {title} {{Extending Hirshfeld-I to bulk and periodic materials}},\ }\href {https://doi.org/10.1002/jcc.23088} {\bibfield  {journal} {\bibinfo  {journal} {J. Comput. Chem.}\ }\textbf {\bibinfo {volume} {34}},\ \bibinfo {pages} {405} (\bibinfo {year} {2013}{\natexlab{a}})}\BibitemShut {NoStop}%
\bibitem [{\citenamefont {Vanpoucke}\ \emph {et~al.}(2013{\natexlab{b}})\citenamefont {Vanpoucke}, \citenamefont {Van~Driessche},\ and\ \citenamefont {Bultinck}}]{VanpouckeDannyEP:2013bJComputChem}%
  \BibitemOpen
  \bibfield  {author} {\bibinfo {author} {\bibfnamefont {D.~E.~P.}\ \bibnamefont {Vanpoucke}}, \bibinfo {author} {\bibfnamefont {I.}~\bibnamefont {Van~Driessche}},\ and\ \bibinfo {author} {\bibfnamefont {P.}~\bibnamefont {Bultinck}},\ }\bibfield  {title} {\bibinfo {title} {Reply to `{C}omment on ``{E}xtending {H}irshfeld-{I} to bulk and periodic materials'' '},\ }\href {https://doi.org/10.1002/jcc.23193} {\bibfield  {journal} {\bibinfo  {journal} {J. Comput. Chem.}\ }\textbf {\bibinfo {volume} {34}},\ \bibinfo {pages} {422} (\bibinfo {year} {2013}{\natexlab{b}})}\BibitemShut {NoStop}%
\bibitem [{\citenamefont {Vanpoucke}(2019)}]{HIVE_REF}%
  \BibitemOpen
  \bibfield  {author} {\bibinfo {author} {\bibfnamefont {D.~E.~P.}\ \bibnamefont {Vanpoucke}},\ } {\bibinfo {title} {{HIVE-tools} v4.x}},\ \bibinfo {howpublished} {\url{https://github.com/DannyVanpoucke/HIVE4-tools}} (\bibinfo {year} {2019})\BibitemShut {NoStop}%
\bibitem [{\citenamefont {Lebedev}\ and\ \citenamefont {Laikov}(1999)}]{LebedevVI_grid:1999DokladyMath}%
  \BibitemOpen
  \bibfield  {author} {\bibinfo {author} {\bibfnamefont {V.~I.}\ \bibnamefont {Lebedev}}\ and\ \bibinfo {author} {\bibfnamefont {D.}~\bibnamefont {Laikov}},\ }\bibfield  {title} {\bibinfo {title} {{Q}uadrature formula for the sphere of $131$-th algebraic order of accuracy},\ } {\bibfield  {journal} {\bibinfo  {journal} {Doklady Mathematics}\ }\textbf {\bibinfo {volume} {59}},\ \bibinfo {pages} {477} (\bibinfo {year} {1999})}\BibitemShut {NoStop}%
\bibitem [{\citenamefont {Sch\"{u}ler}\ \emph {et~al.}(2018)\citenamefont {Sch\"{u}ler}, \citenamefont {Peil}, \citenamefont {Kraberger}, \citenamefont {Pordzik}, \citenamefont {Marsman}, \citenamefont {Kresse}, \citenamefont {Wehling},\ and\ \citenamefont {Aichhorn}}]{SchulerM:2018JPhysCondMat}%
  \BibitemOpen
  \bibfield  {author} {\bibinfo {author} {\bibfnamefont {M.}~\bibnamefont {Sch\"{u}ler}}, \bibinfo {author} {\bibfnamefont {O.~E.}\ \bibnamefont {Peil}}, \bibinfo {author} {\bibfnamefont {G.~J.}\ \bibnamefont {Kraberger}}, \bibinfo {author} {\bibfnamefont {R.}~\bibnamefont {Pordzik}}, \bibinfo {author} {\bibfnamefont {M.}~\bibnamefont {Marsman}}, \bibinfo {author} {\bibfnamefont {G.}~\bibnamefont {Kresse}}, \bibinfo {author} {\bibfnamefont {T.~O.}\ \bibnamefont {Wehling}},\ and\ \bibinfo {author} {\bibfnamefont {M.}~\bibnamefont {Aichhorn}},\ }\bibfield  {title} {\bibinfo {title} {Charge self-consistent many-body corrections using optimized projected localized orbitals},\ }\href {https://doi.org/10.1088/1361-648X/aae80a} {\bibfield  {journal} {\bibinfo  {journal} {J. Phys. Condens. Matter}\ }\textbf {\bibinfo {volume} {30}},\ \bibinfo {pages} {475901} (\bibinfo {year} {2018})}\BibitemShut {NoStop}%
\bibitem [{\citenamefont {Vanpoucke}(2020)}]{VanpoukeDEP:ComputMaterSci2020}%
  \BibitemOpen
  \bibfield  {author} {\bibinfo {author} {\bibfnamefont {D.~E.~P.}\ \bibnamefont {Vanpoucke}},\ }\bibfield  {title} {\bibinfo {title} {Partitioning the vibrational spectrum: {F}ingerprinting defects in solids},\ }\href {https://doi.org/10.1016/j.commatsci.2020.109736} {\bibfield  {journal} {\bibinfo  {journal} {Comput. Mater. Sci.}\ }\textbf {\bibinfo {volume} {181}},\ \bibinfo {pages} {109736} (\bibinfo {year} {2020})}\BibitemShut {NoStop}%
\bibitem [{\citenamefont {Mavrin}(2018)}]{Mavrin:2018JExpTheorPhys}%
  \BibitemOpen
  \bibfield  {author} {\bibinfo {author} {\bibfnamefont {B.~N.}\ \bibnamefont {Mavrin}},\ }\bibfield  {title} {\bibinfo {title} {{Band and Impurity States in Dimond with the (MV)$^-$ (M = Si, Ge, Sn) Centers Based on ab Initio Calculations}},\ }\href {https://doi.org/10.1134/S1063776118120208} {\bibfield  {journal} {\bibinfo  {journal} {J. Exp. Theor. Phys.}\ }\textbf {\bibinfo {volume} {127}},\ \bibinfo {pages} {1016} (\bibinfo {year} {2018})}\BibitemShut {NoStop}%
\bibitem [{fn:({\natexlab{a}})}]{fn:KSindices}%
  \BibitemOpen
  \bibinfo {note} {{I}n theory one could prevent the change exchange of states during the optimization algorithm, though this will have an impact on the convergence being slowed down, and would result in an unordered series of energy levels.}\BibitemShut {Stop}%
\bibitem [{fn:({\natexlab{b}})}]{fn:HSEHSE}%
  \BibitemOpen
  \bibinfo {note} {{T}his is the case for both HSE@PBE and HSE@HSE calculations.}\BibitemShut {Stop}%
\bibitem [{\citenamefont {D'Haenens-Johansson}\ \emph {et~al.}(2011)\citenamefont {D'Haenens-Johansson}, \citenamefont {Edmonds}, \citenamefont {Green}, \citenamefont {Newton}, \citenamefont {Davies}, \citenamefont {Martineau}, \citenamefont {Khan},\ and\ \citenamefont {Twitchen}}]{Haesens:2011PhysRevB_SiV0vsSiVminZPL}%
  \BibitemOpen
  \bibfield  {author} {\bibinfo {author} {\bibfnamefont {U.~F.~S.}\ \bibnamefont {D'Haenens-Johansson}}, \bibinfo {author} {\bibfnamefont {A.~M.}\ \bibnamefont {Edmonds}}, \bibinfo {author} {\bibfnamefont {B.~L.}\ \bibnamefont {Green}}, \bibinfo {author} {\bibfnamefont {M.~E.}\ \bibnamefont {Newton}}, \bibinfo {author} {\bibfnamefont {G.}~\bibnamefont {Davies}}, \bibinfo {author} {\bibfnamefont {P.~M.}\ \bibnamefont {Martineau}}, \bibinfo {author} {\bibfnamefont {R.~U.~A.}\ \bibnamefont {Khan}},\ and\ \bibinfo {author} {\bibfnamefont {D.~J.}\ \bibnamefont {Twitchen}},\ }\bibfield  {title} {\bibinfo {title} {Optical properties of the neutral silicon split-vacancy center in diamond},\ }\href {https://doi.org/10.1103/PhysRevB.84.245208} {\bibfield  {journal} {\bibinfo  {journal} {Phys. Rev. B}\ }\textbf {\bibinfo {volume} {84}},\ \bibinfo {pages} {245208} (\bibinfo {year} {2011})}\BibitemShut {NoStop}%
\bibitem [{\citenamefont {Green}\ \emph {et~al.}(2017)\citenamefont {Green}, \citenamefont {Mottishaw}, \citenamefont {Breeze}, \citenamefont {Edmonds}, \citenamefont {D'Haenens-Johansson}, \citenamefont {Doherty}, \citenamefont {Williams}, \citenamefont {Twitchen},\ and\ \citenamefont {Newton}}]{Green:2017PhysRevLett_SiV0ZPL}%
  \BibitemOpen
  \bibfield  {author} {\bibinfo {author} {\bibfnamefont {B.~L.}\ \bibnamefont {Green}}, \bibinfo {author} {\bibfnamefont {S.}~\bibnamefont {Mottishaw}}, \bibinfo {author} {\bibfnamefont {B.~G.}\ \bibnamefont {Breeze}}, \bibinfo {author} {\bibfnamefont {A.~M.}\ \bibnamefont {Edmonds}}, \bibinfo {author} {\bibfnamefont {U.~F.~S.}\ \bibnamefont {D'Haenens-Johansson}}, \bibinfo {author} {\bibfnamefont {M.~W.}\ \bibnamefont {Doherty}}, \bibinfo {author} {\bibfnamefont {S.~D.}\ \bibnamefont {Williams}}, \bibinfo {author} {\bibfnamefont {D.~J.}\ \bibnamefont {Twitchen}},\ and\ \bibinfo {author} {\bibfnamefont {M.~E.}\ \bibnamefont {Newton}},\ }\bibfield  {title} {\bibinfo {title} {{Neutral Silicon-Vacancy Center in Diamond: Spin Polarization and Lifetimes}},\ }\href {https://doi.org/10.1103/PhysRevLett.119.096402} {\bibfield  {journal} {\bibinfo  {journal} {Phys. Rev. Lett.}\ }\textbf {\bibinfo {volume} {119}},\ \bibinfo {pages} {096402} (\bibinfo {year} {2017})}\BibitemShut {NoStop}%
\bibitem [{\citenamefont {Krivobok}\ \emph {et~al.}(2020)\citenamefont {Krivobok}, \citenamefont {Ekimov}, \citenamefont {Lyapin}, \citenamefont {Nikolaev}, \citenamefont {Skakov}, \citenamefont {Razgulov},\ and\ \citenamefont {Kondrin}}]{KrivobokV:2020PRB}%
  \BibitemOpen
  \bibfield  {author} {\bibinfo {author} {\bibfnamefont {V.~S.}\ \bibnamefont {Krivobok}}, \bibinfo {author} {\bibfnamefont {E.~A.}\ \bibnamefont {Ekimov}}, \bibinfo {author} {\bibfnamefont {S.~G.}\ \bibnamefont {Lyapin}}, \bibinfo {author} {\bibfnamefont {S.~N.}\ \bibnamefont {Nikolaev}}, \bibinfo {author} {\bibfnamefont {Y.~A.}\ \bibnamefont {Skakov}}, \bibinfo {author} {\bibfnamefont {A.~A.}\ \bibnamefont {Razgulov}},\ and\ \bibinfo {author} {\bibfnamefont {M.~V.}\ \bibnamefont {Kondrin}},\ }\bibfield  {title} {\bibinfo {title} {{Observation of a 1.979-eV spectral line of a germanium-related color center in microdiamonds and nanodiamonds}},\ }\href {https://doi.org/10.1103/physrevb.101.144103} {\bibfield  {journal} {\bibinfo  {journal} {Phys. Rev. B.}\ }\textbf {\bibinfo {volume} {101}},\ \bibinfo {pages} {144103} (\bibinfo {year} {2020})}\BibitemShut {NoStop}%
\bibitem [{\citenamefont {Greentree}\ \emph {et~al.}(2008)\citenamefont {Greentree}, \citenamefont {Fairchild}, \citenamefont {Hossain},\ and\ \citenamefont {Prawer}}]{greentree:2008MatToday_NV}%
  \BibitemOpen
  \bibfield  {author} {\bibinfo {author} {\bibfnamefont {A.~D.}\ \bibnamefont {Greentree}}, \bibinfo {author} {\bibfnamefont {B.~A.}\ \bibnamefont {Fairchild}}, \bibinfo {author} {\bibfnamefont {F.~M.}\ \bibnamefont {Hossain}},\ and\ \bibinfo {author} {\bibfnamefont {S.}~\bibnamefont {Prawer}},\ }\bibfield  {title} {\bibinfo {title} {Diamond integrated quantum photonics},\ }\href {https://doi.org/10.1016/S1369-7021(08)70176-7} {\bibfield  {journal} {\bibinfo  {journal} {Mater. Today}\ }\textbf {\bibinfo {volume} {11}},\ \bibinfo {pages} {22} (\bibinfo {year} {2008})}\BibitemShut {NoStop}%
\bibitem [{\citenamefont {L{\"u}hmann}\ \emph {et~al.}(2020)\citenamefont {L{\"u}hmann}, \citenamefont {K{\"u}pper}, \citenamefont {Dietel}, \citenamefont {Staacke}, \citenamefont {Meijer},\ and\ \citenamefont {Pezzagna}}]{Luhman:2020ACSPhoton_ChargeTuneSnV}%
  \BibitemOpen
  \bibfield  {author} {\bibinfo {author} {\bibfnamefont {T.}~\bibnamefont {L{\"u}hmann}}, \bibinfo {author} {\bibfnamefont {J.}~\bibnamefont {K{\"u}pper}}, \bibinfo {author} {\bibfnamefont {S.}~\bibnamefont {Dietel}}, \bibinfo {author} {\bibfnamefont {R.}~\bibnamefont {Staacke}}, \bibinfo {author} {\bibfnamefont {J.}~\bibnamefont {Meijer}},\ and\ \bibinfo {author} {\bibfnamefont {S.}~\bibnamefont {Pezzagna}},\ }\bibfield  {title} {\bibinfo {title} {{Charge-State Tuning of Single SnV Centers in Diamond}},\ }\href {https://doi.org/10.1021/acsphotonics.0c01123} {\bibfield  {journal} {\bibinfo  {journal} {ACS Photon.}\ }\textbf {\bibinfo {volume} {7}},\ \bibinfo {pages} {3376} (\bibinfo {year} {2020})}\BibitemShut {NoStop}%
\bibitem [{\citenamefont {Gutiérrez-Finol}\ \emph {et~al.}(2025)\citenamefont {Gutiérrez-Finol}, \citenamefont {Ullah}, \citenamefont {González-Béjar},\ and\ \citenamefont {Gaita-Ariño}}]{Gutierrez:2025GreenChem_Frugal}%
  \BibitemOpen
  \bibfield  {author} {\bibinfo {author} {\bibfnamefont {G.~M.}\ \bibnamefont {Gutiérrez-Finol}}, \bibinfo {author} {\bibfnamefont {A.}~\bibnamefont {Ullah}}, \bibinfo {author} {\bibfnamefont {M.}~\bibnamefont {González-Béjar}},\ and\ \bibinfo {author} {\bibfnamefont {A.}~\bibnamefont {Gaita-Ariño}},\ }\bibfield  {title} {\bibinfo {title} {A call for frugal modelling: two case studies involving molecular spin dynamics},\ }\href {https://doi.org/10.1039/D4GC04900D} {\bibfield  {journal} {\bibinfo  {journal} {Green Chem.}\ }\textbf {\bibinfo {volume} {27}},\ \bibinfo {pages} {3167} (\bibinfo {year} {2025})}\BibitemShut {NoStop}%
\bibitem [{\citenamefont {Bosoni}\ \emph {et~al.}(2024)\citenamefont {Bosoni}, \citenamefont {Beal}, \citenamefont {Bercx}, \citenamefont {Blaha}, \citenamefont {Bl\"{u}gel}, \citenamefont {Br\"{o}der}, \citenamefont {Callsen}, \citenamefont {Cottenier}, \citenamefont {Degomme}, \citenamefont {Dikan}, \citenamefont {Eimre}, \citenamefont {Flage-Larsen}, \citenamefont {Fornari}, \citenamefont {Garcia}, \citenamefont {Genovese}, \citenamefont {Giantomassi}, \citenamefont {Huber}, \citenamefont {Janssen}, \citenamefont {Kastlunger}, \citenamefont {Krack}, \citenamefont {Kresse}, \citenamefont {K\"{u}hne}, \citenamefont {Lejaeghere}, \citenamefont {Madsen}, \citenamefont {Marsman}, \citenamefont {Marzari}, \citenamefont {Michalicek}, \citenamefont {Mirhosseini}, \citenamefont {M\"{u}ller}, \citenamefont {Petretto}, \citenamefont {Pickard}, \citenamefont {Pon\'{e}}, \citenamefont {Rignanese}, \citenamefont {Rubel}, \citenamefont {Ruh}, \citenamefont {Sluydts}, \citenamefont {Vanpoucke}, \citenamefont {Vijay},
  \citenamefont {Wolloch}, \citenamefont {Wortmann}, \citenamefont {Yakutovich}, \citenamefont {Yu}, \citenamefont {Zadoks}, \citenamefont {Zhu},\ and\ \citenamefont {Pizzi}}]{BosoniE:NatRevPhys2023}%
  \BibitemOpen
  \bibfield  {author} {\bibinfo {author} {\bibfnamefont {E.}~\bibnamefont {Bosoni}}, \bibinfo {author} {\bibfnamefont {L.}~\bibnamefont {Beal}}, \bibinfo {author} {\bibfnamefont {M.}~\bibnamefont {Bercx}}, \bibinfo {author} {\bibfnamefont {P.}~\bibnamefont {Blaha}}, \bibinfo {author} {\bibfnamefont {S.}~\bibnamefont {Bl\"{u}gel}}, \bibinfo {author} {\bibfnamefont {J.}~\bibnamefont {Br\"{o}der}}, \bibinfo {author} {\bibfnamefont {M.}~\bibnamefont {Callsen}}, \bibinfo {author} {\bibfnamefont {S.}~\bibnamefont {Cottenier}}, \bibinfo {author} {\bibfnamefont {A.}~\bibnamefont {Degomme}}, \bibinfo {author} {\bibfnamefont {V.}~\bibnamefont {Dikan}}, \bibinfo {author} {\bibfnamefont {K.}~\bibnamefont {Eimre}}, \bibinfo {author} {\bibfnamefont {E.}~\bibnamefont {Flage-Larsen}}, \bibinfo {author} {\bibfnamefont {M.}~\bibnamefont {Fornari}}, \bibinfo {author} {\bibfnamefont {A.}~\bibnamefont {Garcia}}, \bibinfo {author} {\bibfnamefont {L.}~\bibnamefont {Genovese}}, \bibinfo {author} {\bibfnamefont {M.}~\bibnamefont
  {Giantomassi}}, \bibinfo {author} {\bibfnamefont {S.~P.}\ \bibnamefont {Huber}}, \bibinfo {author} {\bibfnamefont {H.}~\bibnamefont {Janssen}}, \bibinfo {author} {\bibfnamefont {G.}~\bibnamefont {Kastlunger}}, \bibinfo {author} {\bibfnamefont {M.}~\bibnamefont {Krack}}, \bibinfo {author} {\bibfnamefont {G.}~\bibnamefont {Kresse}}, \bibinfo {author} {\bibfnamefont {T.~D.}\ \bibnamefont {K\"{u}hne}}, \bibinfo {author} {\bibfnamefont {K.}~\bibnamefont {Lejaeghere}}, \bibinfo {author} {\bibfnamefont {G.~K.~H.}\ \bibnamefont {Madsen}}, \bibinfo {author} {\bibfnamefont {M.}~\bibnamefont {Marsman}}, \bibinfo {author} {\bibfnamefont {N.}~\bibnamefont {Marzari}}, \bibinfo {author} {\bibfnamefont {G.}~\bibnamefont {Michalicek}}, \bibinfo {author} {\bibfnamefont {H.}~\bibnamefont {Mirhosseini}}, \bibinfo {author} {\bibfnamefont {T.~M.~A.}\ \bibnamefont {M\"{u}ller}}, \bibinfo {author} {\bibfnamefont {G.}~\bibnamefont {Petretto}}, \bibinfo {author} {\bibfnamefont {C.~J.}\ \bibnamefont {Pickard}}, \bibinfo {author}
  {\bibfnamefont {S.}~\bibnamefont {Pon\'{e}}}, \bibinfo {author} {\bibfnamefont {G.-M.}\ \bibnamefont {Rignanese}}, \bibinfo {author} {\bibfnamefont {O.}~\bibnamefont {Rubel}}, \bibinfo {author} {\bibfnamefont {T.}~\bibnamefont {Ruh}}, \bibinfo {author} {\bibfnamefont {M.}~\bibnamefont {Sluydts}}, \bibinfo {author} {\bibfnamefont {D.~E.~P.}\ \bibnamefont {Vanpoucke}}, \bibinfo {author} {\bibfnamefont {S.}~\bibnamefont {Vijay}}, \bibinfo {author} {\bibfnamefont {M.}~\bibnamefont {Wolloch}}, \bibinfo {author} {\bibfnamefont {D.}~\bibnamefont {Wortmann}}, \bibinfo {author} {\bibfnamefont {A.~V.}\ \bibnamefont {Yakutovich}}, \bibinfo {author} {\bibfnamefont {J.}~\bibnamefont {Yu}}, \bibinfo {author} {\bibfnamefont {A.}~\bibnamefont {Zadoks}}, \bibinfo {author} {\bibfnamefont {B.}~\bibnamefont {Zhu}},\ and\ \bibinfo {author} {\bibfnamefont {G.}~\bibnamefont {Pizzi}},\ }\bibfield  {title} {\bibinfo {title} {How to verify the precision of density-functional-theory implementations via reproducible and universal
  workflows},\ }\href {https://doi.org/10.1038/s42254-023-00655-3} {\bibfield  {journal} {\bibinfo  {journal} {Nat. Rev. Phys.}\ }\textbf {\bibinfo {volume} {6}},\ \bibinfo {pages} {45} (\bibinfo {year} {2024})}\BibitemShut {NoStop}%
\bibitem [{\citenamefont {Perdew}\ and\ \citenamefont {Schmidt}(2001)}]{PerdewJacobs:AIPConfProc2001}%
  \BibitemOpen
  \bibfield  {author} {\bibinfo {author} {\bibfnamefont {J.~P.}\ \bibnamefont {Perdew}}\ and\ \bibinfo {author} {\bibfnamefont {K.}~\bibnamefont {Schmidt}},\ }\bibfield  {title} {\bibinfo {title} {{Jacob’s ladder of density functional approximations for the exchange-correlation energy}},\ }\href {https://doi.org/10.1063/1.1390175} {\bibfield  {journal} {\bibinfo  {journal} {AIP Conference Proceedings}\ }\textbf {\bibinfo {volume} {577}},\ \bibinfo {pages} {1} (\bibinfo {year} {2001})}\BibitemShut {NoStop}%
\bibitem [{fn:({\natexlab{c}})}]{fn:CO2}%
  \BibitemOpen
  \bibinfo {note} {{T}he carbon footprint is estimated based on the carbon footprint of the energy production in Belgium (132.672 g/kWh \cite{owid-co2-belgium}), the powerusage of the AMD Epyc 7H12 and 7763 64-core CPUs used in the compute cluster (280W), and the fact that CPUh reflect CPU-core-hours.}\BibitemShut {Stop}%
\bibitem [{\citenamefont {Ritchie}\ and\ \citenamefont {Rosado}(2025)}]{owid-co2-belgium}%
  \BibitemOpen
  \bibfield  {author} {\bibinfo {author} {\bibfnamefont {H.}~\bibnamefont {Ritchie}}\ and\ \bibinfo {author} {\bibfnamefont {P.}~\bibnamefont {Rosado}},\ }\bibfield  {title} {\bibinfo {title} {CO$_2$ and greenhouse gas emissions},\ } {\bibfield  {journal} {\bibinfo  {journal} {Our World in Data}\ } (\bibinfo {year} {2025})},\ \bibinfo {note} {https://ourworldindata.org/profile/co2/belgium}\BibitemShut {NoStop}%
\end{thebibliography}


\clearpage




\setcounter{figure}{0}
\setcounter{table}{0}
\renewcommand{\thefigure}{SI.\arabic{figure}}
\renewcommand{\thetable}{SI.\Roman{table}}
\renewcommand{\thesection}{SI.\arabic{section}}
\renewcommand{\thesubsection}{\thesection.\arabic{subsection}}

\onecolumngrid
\widetext
\begin{center}
\textbf{\large Supporting information for ``Modeling the Zero Phonon Line of strained SnV centers in diamond''\\}
\text{Danny E. P. Vanpoucke$^{1,2}$}\\
\textit{$^{1}$UHasselt, Institute for Materials Research (IUMAT), Quantum \& Artificial inTelligence design Of Materials (QuATOMs), Martelarenlaan 42, B-3500 Hasselt, Belgium\\}
\textit{$^{2}$imec, IUMAT, Wetenschapspark 1, B-3590 Diepenbeek, Belgium\\}
\end{center}

\tableofcontents
\newpage

\section{Supplemental Theoretical Data}
Several different configurations of the occupancies of the spin-minority channel Kohn--Sham(KS) states close to the Valence Band Maximum(VBM) are considered in this work. With the aim of having a clear and simple notation to identify these configurations, a seven digit string is used to indicate the occupancy of the top five bands below the VBM (\textit{i.e.}, $2\times e_u$, $2\times$VBM-1, and VBM) and the two gap states ($2\times e_g$). As the order of the states is a point of uncertainty, the string will only represent the index position position of the KS state. The actual character of the states will be indicated in a different manner. An empty or filled KS state is indicated as $0$ or $1$ respectively, while a half filled KS state (for example as a result of equal filling of two degenerate KS states with a single electron) is indicated as $5$. Within this notation the ground state configuration of the SnV$^0$ would be represented as $1111100$, while the ground state of the SnV$^-$ would be represented as $1111110$ or $1111155$ indicating a single electron in a single $e_g$ state or the electron distributed over the degenerate $e_g$ states.\\
To gain some insights into the redistribution of the electron density, and it's convergence with regard to the supercell size, the Hirshfeld-I charges for all atoms in the systems are calculated in the ground state configurations. The Sn atoms receives a large positive charge, while the six C atoms bound to it have a significant negative charge. In case of the negative charge state, when one extra electron is added to the system, the extra electron clearly localizes on the color center, as can be seen in Table~\ref{SI:table:HI}.\\

\begin{table}[!h]
    \caption{The calculated Hirshfeld-I charges associated with the SnV color center the neutral and negative charge state. The atomic charges are calculated using supercells of $512$ ($0.195$\%) and $1000$ ($0.100$\%) atoms (color center concentration) in their equilibrium volume. Charges are given in units of $e$.}\label{SI:table:HI}
    \begin{ruledtabular}
    \begin{tabular}{l|rrrr}
         & \multicolumn{2}{c}{$0.195$\%} & \multicolumn{2}{c}{$0.100$\%} \\ 
         & SnV$^0$ & SnV$^-$ & SnV$^0$ & SnV$^-$ \\
         \hline
         $Q_{Sn}$  &  $2.187$ &  $2.241$ &  $2.189$ &  $2.248$  \\
         $Q_{C}$   & $-0.691$ & $-0.796$ & $-0.689$ & $-0.796$  \\
         $Q_{SnV}$ & $-1.958$ & $-2.537$ & $-1.947$ & $-2.527$  \\
    \end{tabular}
    \end{ruledtabular}
\end{table}

\newpage
\subsection{The ground state configurations}
\begin{table}[!h]
    \caption{Position and color center character of the spin-minority channel KS states around the VBM for SnV$^0$ and SnV$^-$ using a $512$ atom conventional supercell. The Conduction Band Minimum is included as KS state with index $1029$. The VBM (index $1026$) and $e_g$ states (indices $1027$ and $1028$) retain their order.}\label{SI:table:GSconfigs444}
    \begin{ruledtabular}
    \begin{tabular}{l|rrrrrr}
              & \multicolumn{3}{c}{PBE} & \multicolumn{3}{c}{HSE} \\
              & SnV$^0$ & SnV$^-$ & SnV$^-$ & SnV$^0$ & SnV$^-$ & SnV$^-$ \\
              & 1111100 & 1111110 & 1111155 &  1111100 & 1111110 & 1111155 \\
        index &  \multicolumn{6}{c}{KS energies (eV)} \\
         \hline
         $1022$ &  $12.143$ & $12.419$ & $12.419$  &  $11.980$ &  $12.145$ &  $12.176$ \\
         $1023$ &  $12.143$ & $12.419$ & $12.419$  &  $11.980$ &  $12.170$ &  $12.176$ \\
         $1024$ &  $12.185$ & $12.436$ & $12.437$  &  $11.217$ &  $12.182$ &  $12.184$ \\
         $1025$ &  $12.185$ & $12.436$ & $12.437$  &  $11.217$ &  $12.221$ &  $12.185$ \\
         $1026$ &  $12.543$ & $12.497$ & $12.498$  &  $11.271$ &  $12.241$ &  $12.241$ \\
         $1027$ &  $14.001$ & $14.347$ & $14.348$  &  $14.533$ &  $14.312$ &  $14.556$ \\
         $1028$ &  $14.001$ & $14.347$ & $14.348$  &  $14.533$ &  $14.808$ &  $14.558$ \\
         $1029$ &  $16.686$ & $16.621$ & $16.621$  &  $17.635$ &  $17.588$ &  $17.589$ \\
         \hline
    index &  \multicolumn{6}{c}{cumulative color center character} \\
         \hline
         $1022$ &  $0.296$ & $0.385$ & $0.385$  &  $0.355$ &  $0.390$ &  $0.013$ \\
         $1023$ &  $0.296$ & $0.386$ & $0.386$  &  $0.355$ &  $0.017$ &  $0.013$ \\
         $1024$ &  $0.018$ & $0.013$ & $0.013$  &  $0.015$ &  $0.010$ &  $0.396$ \\
         $1025$ &  $0.018$ & $0.017$ & $0.017$  &  $0.013$ &  $0.403$ &  $0.396$ \\
         $1026$ &  $0.007$ & $0.007$ & $0.007$  &  $0.007$ &  $0.007$ &  $0.007$ \\
         $1027$ &  $0.509$ & $0.528$ & $0.527$  &  $0.538$ &  $0.538$ &  $0.543$ \\
         $1028$ &  $0.511$ & $0.528$ & $0.528$  &  $0.536$ &  $0.547$ &  $0.540$ \\
         $1029$ &  $0.010$ & $0.004$ & $0.004$  &  $0.008$ &  $0.006$ &  $0.006$ \\
    \end{tabular}
    \end{ruledtabular}
\end{table}

\begin{table}[!h]
    \caption{Position and color center character of the spin-minority channel KS states around the VBM for SnV$^0$ and SnV$^-$ using a $1000$ atom conventional supercell. The Conduction Band Minimum is included as KS state with index $2005$. The $e_g$ states are found at indices $2003$ and $2004$, while the $e_u$ state indices vary per functional and charge state.}\label{SI:table:GSconfigs555}
    \begin{ruledtabular}
    \begin{tabular}{l|rrrrrr}
              & \multicolumn{2}{c}{PBE} & \multicolumn{2}{c}{HSE} \\
              & SnV$^0$ & SnV$^-$ & SnV$^0$ & SnV$^-$ \\
              & 1111100 & 1111110 &  1111100 & 1111110  \\
        index &  \multicolumn{4}{c}{KS energies (eV)} \\
         \hline
         $1998$ &  $12.236$ & $12.530$ & $12.044$  &  $12.257$  \\
         $1999$ &  $12.236$ & $12.530$ & $12.044$  &  $12.264$  \\
         $2000$ &  $12.558$ & $12.555$ & $12.295$  &  $12.278$  \\
         $2001$ &  $12.558$ & $12.555$ & $12.295$  &  $12.296$  \\
         $2002$ &  $12.585$ & $12.564$ & $12.320$  &  $12.348$  \\
         $2003$ &  $14.004$ & $14.461$ & $14.548$  &  $14.420$  \\
         $2004$ &  $14.004$ & $14.461$ & $14.548$  &  $14.916$  \\
         $2005$ &  $16.754$ & $16.705$ & $17.702$  &  $17.653$  \\
         \hline
    index &  \multicolumn{4}{c}{cumulative color center character} \\
         \hline
         $1998$ &  $0.170$ & $0.009$ & $0.295$  &  $0.009$  \\
         $1999$ &  $0.168$ & $0.005$ & $0.295$  &  $0.007$  \\
         $2000$ &  $0.010$ & $0.373$ & $0.005$  &  $0.378$  \\
         $2001$ &  $0.010$ & $0.373$ & $0.005$  &  $0.007$  \\
         $2002$ &  $0.000$ & $0.007$ & $0.000$  &  $0.394$  \\
         $2003$ &  $0.514$ & $0.531$ & $0.539$  &  $0.541$  \\
         $2004$ &  $0.514$ & $0.527$ & $0.539$  &  $0.548$  \\
         $2005$ &  $0.003$ & $0.000$ & $0.004$  &  $0.000$  \\
    \end{tabular}
    \end{ruledtabular}
\end{table}

\begin{table}[!t]
    \caption{The position of the ZPL in the $\Delta$KS model for the SnV center using the PBE and HSE functionals and supercells of $512$ ($0.195$\%) and $1000$ ($0.100$\%) atoms (color center concentration). ZPL position is given in eV units, and the experimentally observed peak position is considered to be at $1.9997$ eV.}\label{SI:table:ZPLeq_DKS}
    \begin{ruledtabular}
    \begin{tabular}{ll|rrr}
         & functional & $0.195$\% & $0.100$\% & dilute limit \\
         \hline
         SnV$^0$ & PBE     & $1.858$ & $1.768$ & $1.674$  \\
         SnV$^0$ & HSE@PBE & $2.497$ & $2.452$ & $2.406$  \\
         SnV$^0$ & HSE@HSE & $2.553$ & $2.504$ & $2.453$  \\
         SnV$^-$ & PBE     & $1.928$ & $1.906$ & $1.882$  \\
         SnV$^-$ & HSE@PBE & $2.525$ & $2.506$ & $2.487$  \\
         SnV$^-$ & HSE@HSE & $2.587$ & $2.568$ & $2.549$  \\
    \end{tabular}
    \end{ruledtabular}
\end{table}

\newpage
\subsection{Role of the electronic configuration in the Franck--Codon model.}
\begin{table}[!h]
\caption{Position and color center character of the spin-minority channel KS states for various excited state configurations of the SnV$^0$ at the different stages (indicated, \textit{cf.} Fig.~2 of the manuscript) in the Franck--Codon approximation. The KS state from which an electron is/was excited is indicated in bold in each sub-table. For the KS state positions the states with $e_u$ character are indicated in pink, while the $e_g$ states are indicated in blue and the conduction band minimum (CBM) is indicated in orange. For the KS state characters, the states with high color center character (fraction $\in [0..1]$) are highlighted in green. The values of the ground state configuration are provided for comparison.}
    \label{SI:figTable:TabSSnV0_PBE_FC}
    \begin{center}
    \includegraphics[width=0.75\linewidth]{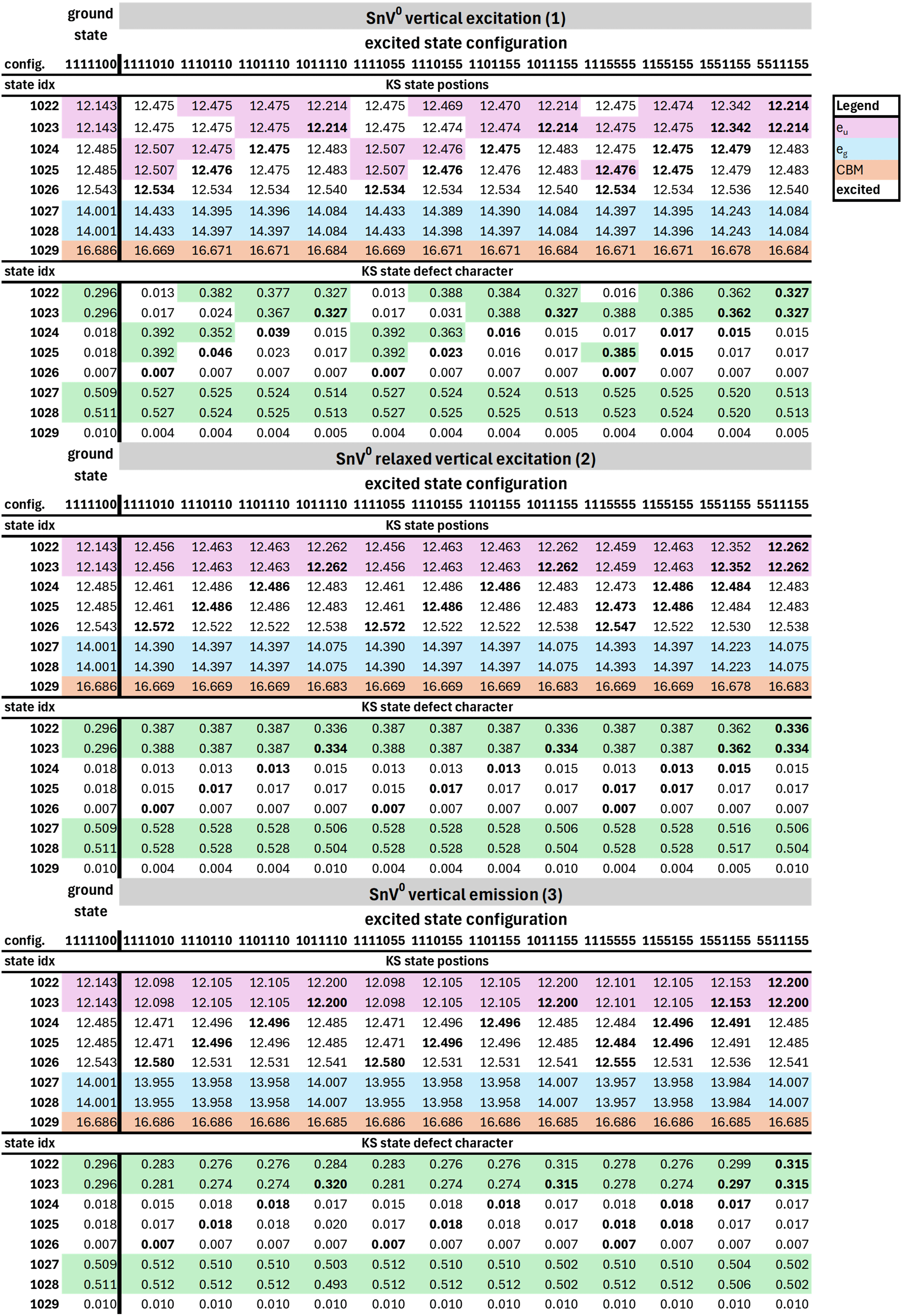}
    \end{center}
\end{table}

\begin{table}[!h]
\caption{Position and color center character of the spin-minority channel KS states for various excited state configurations of the SnV$^-$ at the different stages (indicated, \textit{cf.} Fig.~2 of the manuscript) in the Franck--Codon approximation. The KS state from which an electron is/was excited is indicated in bold in each sub-table. For the KS state positions the states with $e_u$ character are indicated in pink, while the $e_g$ states are indicated in blue and the conduction band minimum (CBM) is indicated in orange. For the KS state characters, the states with high color center character (fraction $\in [0..1]$) are highlighted in green. For the stage after emission (3) both the ground state configuration with the $e_g$ electron degenerate over both KS states and localized in a single $e_g$ state are considered. The values of the ground state configuration are provided for comparison.}
    \label{SI:figTable:TabSSnV0_PBE_FC}
    \begin{center}
    \includegraphics[width=0.6\linewidth]{SITab2_SnVm1_PBE_FC.png}
    \end{center}
\end{table}    

\newpage
\begin{table}[!tb]
    \caption{The calculated absorption, emission and ZPL energies of the SnV$^0$ color center in a $512$ atom supercell using the PBE functional for different excitation configurations. The maximal remaining forces on a single atom (Max. Force) at the three different stages of the Franck--Codon approximation, as well as the calculated Stokes (S) and Anti-Stokes (AS) shifts. For each configuration, the character of the KS state from which the electron was excited (Char.Ext.) is indicated as well.}\label{SI:table:FC_SnV0_configs}
    \begin{ruledtabular}
    \begin{tabular}{l|rrrrrrrrrrrr}
         Config. & \multicolumn{3}{c}{Max. Force [meV/\AA]} & \multicolumn{2}{c}{absorption} & \multicolumn{2}{c}{ZPL} & \multicolumn{2}{c}{emission} & S & AS & Char. Ext. \\
                 & (1) & (2) & (3) & [nm] & [eV] & [nm] & [eV] & [nm] & [eV] & [meV] & [meV] & \\
         \hline
        1111010 & $358.23$ & $0.05$ & $361.33$ & $739.5$ & $1.677$ & $758.0$ & $1.636$ & $777.6$ & $1.594$ & $41$ & $41$ & VBM \\
        1110110 & $579.30$ & $0.16$ & $312.24$ & $715.6$ & $1.733$ & $725.4$ & $1.709$ & $736.7$ & $1.683$ & $23$ & $26$ & VBM-1 \\
        1101110 & $694.04$ & $0.15$ & $312.35$ & $715.1$ & $1.734$ & $725.4$ & $1.709$ & $736.7$ & $1.683$ & $24$ & $26$ & VBM-1 \\
        1011110 & $494.25$ & $0.07$ & $501.83$ & $664.8$ & $1.865$ & $674.5$ & $1.838$ & $684.8$ & $1.811$ & $27$ & $28$ & $e_u$ \\
        1111055 & $358.28$ & $0.05$ & $361.35$ & $739.5$ & $1.677$ & $758.0$ & $1.636$ & $777.6$ & $1.594$ & $41$ & $41$ & VBM \\
        1110155 & $335.79$ & $0.17$ & $312.25$ & $715.8$ & $1.732$ & $725.4$ & $1.709$ & $736.7$ & $1.683$ & $23$ & $26$ & VBM-1 \\
        1101155 & $341.37$ & $0.10$ & $312.24$ & $715.1$ & $1.734$ & $725.4$ & $1.709$ & $736.7$ & $1.683$ & $24$ & $26$ & VBM-1 \\
        1011155 & $494.25$ & $0.07$ & $501.81$ & $664.8$ & $1.865$ & $674.5$ & $1.838$ & $684.8$ & $1.811$ & $27$ & $28$ & $e_u$ \\
        1115555 & $687.53$ & $0.11$ & $334.80$ & $727.4$ & $1.704$ & $737.2$ & $1.682$ & $748.2$ & $1.657$ & $23$ & $25$ & mix \\
        1155155 & $316.04$ & $0.17$ & $312.25$ & $715.1$ & $1.734$ & $725.4$ & $1.709$ & $736.7$ & $1.683$ & $24$ & $26$ & VBM-1 \\
        1551155 & $217.37$ & $0.05$ & $213.02$ & $703.7$ & $1.762$ & $708.9$ & $1.749$ & $714.3$ & $1.736$ & $13$ & $13$ & mix \\
        5511155 & $494.25$ & $0.08$ & $501.81$ & $664.8$ & $1.865$ & $674.5$ & $1.838$ & $684.8$ & $1.811$ & $27$ & $28$ & $e_u$ \\
    \end{tabular}
    \end{ruledtabular}
\end{table}

\begin{table}[!tb]
    \caption{The calculated absorption, emission and ZPL energies of the SnV$^-$ color center in a $512$ atom supercell using the PBE functional for different excitation configurations. The maximal remaining forces on a single atom (Max. Force) at the three different stages of the Franck--Codon approximation, as well as the calculated Stokes (S) and Anti-Stokes (AS) shifts. For each configuration, the character of the KS state from which the electron was excited (Char.Ext.) is indicated as well. The ground state geometry started from and fluoresced to contains the $e_g$ electron in a single KS state (\textit{i.e.}, the 1111110 configuration) }\label{SI:table:FC_SnVm1_configs}
    \begin{ruledtabular}
    \begin{tabular}{l|rrrrrrrrrrrr}
         Config. & \multicolumn{3}{c}{Max. Force [meV/\AA]} & \multicolumn{2}{c}{absorption} & \multicolumn{2}{c}{ZPL} & \multicolumn{2}{c}{emission} & S & AS & Char. Ext. \\
                 & (1) & (2) & (3) & [nm] & [eV] & [nm] & [eV] & [nm] & [eV] & [meV] & [meV] & \\
         \hline
        1111011 & $1178.92$ & $0.05$ & $790.21$ & $644.2$ & $1.925$ & $656.8$ & $1.888$ & $670.5$ & $1.849$ & $37$ & $39$ & $e_u$ \\
        1110111 & $1413.69$ & $0.12$ & $784.65$ & $640.1$ & $1.937$ & $656.8$ & $1.888$ & $670.5$ & $1.849$ & $49$ & $39$ & $e_u$ \\
        1101111 &  $372.03$ & $0.06$ & $735.94$ & $595.8$ & $2.081$ & $607.4$ & $2.041$ & $619.6$ & $2.001$ & $40$ & $40$ & VBM \\
        1011111 &  $323.71$ & $0.14$ & $600.58$ & $576.9$ & $2.149$ & $583.6$ & $2.125$ & $590.4$ & $2.100$ & $25$ & $25$ & VBM-1 \\
        0111111 &  $326.32$ & $0.05$ & $589.93$ & $576.9$ & $2.149$ & $583.6$ & $2.125$ & $590.4$ & $2.100$ & $25$ & $25$ & VBM-1 \\
        1115511 &  $659.11$ & $0.06$ & $780.06$ & $644.1$ & $1.925$ & $656.8$ & $1.888$ & $670.5$ & $1.849$ & $37$ & $39$ & $e_u$ \\
        1155111 &  $556.98$ & $0.08$ & $561.72$ & $632.4$ & $1.961$ & $640.6$ & $1.935$ & $647.2$ & $1.916$ & $25$ & $20$ & mix \\
        1551111 &  $341.45$ & $0.06$ & $702.51$ & $586.2$ & $2.115$ & $592.6$ & $2.092$ & $599.1$ & $2.070$ & $23$ & $23$ & mix \\
        5511111 &  $302.34$ & $0.05$ & $670.06$ & $576.9$ & $2.149$ & $583.6$ & $2.125$ & $590.5$ & $2.100$ & $25$ & $25$ & VBM-1 \\
    \end{tabular}
    \end{ruledtabular}
\end{table}

\subsection{Impact of the k-point set.}
\begin{table}[!h]
    \caption{The position of the ZPL in the Franck--Codon model of the SnV center using the PBE functional and a $512$($0.195$\%) or $1000$($0.100$\%) atom (color center concentration) supercell as function of the k-point set. ZPL position is given in eV units, and the experimentally observed peak position is considered to be at $1.9997$ eV.}\label{SI:table:ZPLeq_FCkpoint_eV}
    \begin{ruledtabular}
    \begin{tabular}{ll|rrr}
          & supercell & $\Gamma$ & $2\times2\times2$ & $3\times3\times3$  \\
         \hline
         SnV$^0$ & $512$ & $1.838$ & $1.708$ & $1.738$  \\
         SnV$^-$ & $512$ & $1.888$ & $1.837$ & $1.840$  \\
         \hline
         SnV$^0$ & $1000$ & $1.787$ & $1.712$ & $1.731$  \\
         SnV$^-$ & $1000$ & $1.867$ & $1.849$ & $1.849$  \\
    \end{tabular}
    \end{ruledtabular}
\end{table}

\clearpage

\subsection{Pressure coefficients under Hydrostatic strain}
\begin{table}[!h]
    \caption{The ZPL pressure coefficients using the $\Delta$KS approximation using the PBE and HSE06 functionals and supercells of $512$ ($0.195$\%) and $1000$ ($0.100$\%) atoms (color center concentration). The pressure coefficient is in meV/GPa units.}\label{SI:table:ZPLshift_dKZ_meVGPa}
    \begin{ruledtabular}
    \begin{tabular}{lll|rrr}
         & functional & k-points & $0.195$\% & $0.100$\% & dilute limit \\
         \hline
         SnV$^0$ & PBE & $\Gamma$           & $3.74$ & $3.21$ & $2.62$  \\
         SnV$^0$ & PBE & $2\times2\times2$  & $3.75$ & $3.17$ & $2.59$  \\
         SnV$^-$ & PBE & $\Gamma$           & $4.11$ & $3.95$ & $3.83$  \\
         SnV$^-$ & PBE & $2\times2\times2$  & $4.08$ & $3.93$ & $3.80$  \\
         SnV$^0$ & HSE@PBE & $\Gamma$       & $4.39$ & $4.12$ & $3.81$  \\
         SnV$^-$ & HSE@PBE & $\Gamma$       & $4.67$ & $4.51$ & $4.39$  \\
    \end{tabular}
    \end{ruledtabular}
\end{table}

\begin{table}[!h]
    \caption{The ZPL pressure coefficients using the Franck--Codon approximation using the PBE and HSE functionals and supercells of $512$ ($0.195$\%) and $1000$ ($0.100$\%) atoms (color center concentration). The pressure coefficient is in meV/GPa units.}\label{SI:table:ZPLshift_FC_meVGPa}
    \begin{ruledtabular}
    \begin{tabular}{lll|rrr}
         & functional & k-points & $0.195$\% & $0.100$\% & dilute limit \\
         \hline
         SnV$^0$ & PBE & $\Gamma$           & $3.82$ & $3.57$ & $3.29$  \\
         SnV$^0$ & PBE & $2\times2\times2$  & $3.28$ & $3.17$ & $3.08$  \\
         SnV$^-$ & PBE & $\Gamma$           & $4.06$ & $3.93$ & $3.83$  \\
         SnV$^-$ & PBE & $2\times2\times2$  & $3.82$ & $3.83$ & $3.87$  \\
         SnV$^-$ & HSE@PBE &$\Gamma$        & $5.00$ & $4.81$ & $4.67$  \\
    \end{tabular}
    \end{ruledtabular}
\end{table}

\end{document}